\documentclass[preprint,12pt,eqsecnum,nofootinbib,amsmath,amssymb]{revtex4}

\newcommand{\datechange}{23 January 2017}

\newcommand{\mytitle}{The Planar Sandwich and Other 1D 
Planar Heat Flow Test Problems in ExactPack}
\usepackage{graphicx} 
\usepackage{dcolumn}  
\usepackage{bm}          
\usepackage{latexsym} 
\usepackage{fancyhdr} 
\usepackage{wrapfig}

\newcommand{\smNH}{{\rm\scriptscriptstyle NH}}
\newcommand{\smL}{{\rm\scriptscriptstyle L}}

\newcommand{\sm}[1]{{\scriptstyle #1}}

\newcommand{\smR}{{\rm\scriptscriptstyle R}}

\newcommand{\bodyskip}{\baselineskip 18pt plus 1pt minus 1pt}
\newcommand{\bibskip}{\baselineskip16pt plus 1pt minus 1pt}
\newcommand{\tableofcontentsskip}{\baselineskip 14pt plus 1pt minus 1pt}
\newcommand{\footnoteskip}{\baselineskip 12pt plus 1pt minus 1pt}
\newcommand{\abstractskip}{\baselineskip 13pt plus 1pt minus 1pt}
\newcommand{\titleskip}{\baselineskip 18pt plus 1pt minus 1pt}
\newcommand{\affiliationskip}{\baselineskip 15pt plus 1pt minus 1pt}

\def\frame#1#2#3#4{\vbox{\hrule height #1pt    
  \hbox{\vrule width #1pt\kern #2pt                     
  \vbox{\kern #2pt                                               
  \vbox{\hsize #3\noindent #4}                            
  \kern #2pt}                                                        
  \kern #2pt\vrule width #1pt}                              
  \hrule height0pt depth #1pt}                            
}
\def\myframe#1{\vbox{\hrule height 0.1pt    
  \hbox{\vrule width 0.1pt\kern 2pt                     
  \vbox{\kern 2pt                                               
  \vbox{\hsize 16.5cm\noindent #1}                            
  \kern 2pt}                                                        
  \kern 2pt\vrule width 0.1pt}                              
  \hrule height0pt depth 0.1pt}                            
}

\def\fitframe #1#2#3{\vbox{\hrule height#1pt  
  \hbox{\vrule width#1pt\kern #2pt             
  \vbox{\kern #2pt\hbox{#3}\kern #2pt}         
  \kern #2pt\vrule width#1pt}                  
  \hrule height0pt depth#1pt}                  
}
\def\shframe #1#2#3#4{\vbox{\hrule height 0pt 
 \hbox{\vrule width #1pt\kern 0pt             
 \vbox{\kern-#1pt\frame{.3}{#2}{#3}{#4}       
 \kern-.3pt}                                  
 \kern-#2pt\vrule width 0pt}                  
 \hrule height #1pt}                          
}

\begin{document}

\preprint{LA-UR-17-20460}

\title{\titleskip
  \mytitle
}

\author{Robert L Singleton Jr}

\vskip 0.2cm 
\affiliation{\affiliationskip
     Los Alamos National Laboratory\\
     Los Alamos, New Mexico 87545, USA
}

\date{\datechange}

\begin{abstract}
\abstractskip
\vskip0.3cm 

\noindent
This report documents the implementation of several related 1D heat 
flow problems in the verification package ExactPack\,\cite{exactpack2014}. 
In particular, the planar sandwich class defined in Ref.~\cite{Dawes2016}, 
as well as the classes PlanarSandwichHot, PlanarSandwichHalf, and other 
generalizations of the planar sandwich problem, are defined and documented 
here. A rather general treatment of 1D heat flow is presented, whose main
results have been implemented in the class Rod1D. All planar sandwich 
classes are derived from the parent class Rod1D.

\end{abstract}

\maketitle

%
%

\pagebreak
%
\tableofcontentsskip
\tableofcontents
\thispagestyle{empty}

\pagebreak
\bodyskip

\newpage
\bodyskip

%

\pagebreak

\section{1D Planar Heat Flow in ExactPack}

\subsection{Use of ExactPack Solvers}

This report documents the implementation of a number of planar 1D 
heat flow problems in the verification package ExactPack\,\cite{exactpack2014}.
The first problem that we consider is the planar sandwich of 
Ref.~\cite{Dawes2016}, and some generalizations thereof, under the 
class names 
\vskip0.4cm 
\begin{enumerate}
  \baselineskip 10pt plus 1pt minus 1pt
  \setlength{\itemsep}{3pt} 
  \setlength{\parskip}{1pt} %
  \setlength{\parsep}{0pt}  %
\item[-] PlanarSandwich
\item[-] PlanarSandwichHot
\item[-] PlanarSandwichHalf
\item[-] Rod1D  .
\end{enumerate}

\noindent
We will describe each of these classes in this section, and  will provide instructions
on how to use them in a python script (for plotting or data  analysis, for example).  
We also provide a pedagogical treatment of 1D heat flow and a detailed derivation 
of the cases treated herein. We have implemented the general 1D heat flow problem
as the class Rod1D, and the planar sandwich classes inherit from this base class.
These classes can be imported and accessed in a python script as follows,

\footnoteskip
\begin{verbatim}
from exactpack.solvers.heat import PlanarSandwich
from exactpack.solvers.heat import PlanarSandwichHot
from exactpack.solvers.heat import PlanarSandwichHalf
from exactpack.solvers.heat import Rod1D  .
\end{verbatim}
\bodyskip

\noindent
To instantiate and use these classes for plotting or analysis, 
one must create a corresponding {\em solver} object; 
for example, an instance of the planar sandwich is created by

\vskip0.2cm
\footnoteskip
\begin{verbatim}
solver = PlanarSandwich(T1=1, T2=0, L=2)  .
\end{verbatim}
\bodyskip

\noindent
This creates an ExactPack solver object  called \lq\lq{}solver\rq\rq{}, with
boundary conditions $T_1=1$ and $T_2=0$, and length $L=2$. All other 
variables take their default values. The solver object does not know anything
about the spatial grid of the solution,  and we must pass an array of 
$x$-values along the length of the rod, as well as a time variable $t$ at 
which to evaluate the solution; for example, 
\vskip0.2cm
\footnoteskip
\begin{verbatim}
x = numpy.linspace(0, 2, 1000)
t = 0.2

soln = solver(x, t)
soln.plot('temperature')  . 
\end{verbatim}
\bodyskip

\noindent
This creates an ExactPack {\em solution} object called \lq\lq{}soln\rq\rq{}.
Solution objects in ExactPack come equipped with a plotting method, as
illustrated in the last line above, in addition to various analysis methods
not shown here. Now that we have reviewed the mechanics of importing and 
using the various planar classes, let us turn to the physics of 1D heat flow.

\subsection{The General 1D Heat Conducting Rod}
\label{sec:generalheatflow}

The planar sandwich is a special case of the simplest form of heat 
conduction problem, namely, 1D heat flow in a rod of length $L$ and 
constant heat conduction $\kappa$. The heat flow equation, along 
with the boundary conditions and an initial condition, take the 
form\,\cite{Berg1966},
\begin{eqnarray}
  {\rm DE}: 
  \hskip3.73cm
  \frac{\partial T(x,t)}{\partial t}
  &=&
  \kappa\, \frac{\partial^2 T(x,t)}{\partial x^2}
  \hskip1.2cm 
  0 < x < L ~{\rm and}~ t > 0
\label{eq_oneDrodAnh}
\\[5pt]
  {\rm BC}:  
  \hskip1.16cm
  \alpha_1 T(0,t) + \beta_1 \partial_x T(0,t) &=& \gamma_1
  \hskip2.8cm t > 0
\label{eq_oneDrodBnhA}    
\\[-3pt]
  \alpha_2 T(L,t) + \beta_2 \partial_x T(L,t) &=& \gamma_2
\label{eq_oneDrodBnhB}    
\\[5pt]
  {\rm IC}:  
  \hskip4.05cm
  T(x,0) &=& T_0(x)   
  \hskip2.2cm 
  0 < x < L
  \ .
\label{eq_oneDrodCnh}
\end{eqnarray}
We use an arbitrary but consistent set of temperature units throughout.
Equation~(\ref{eq_oneDrodAnh}) is the diffusion equation (DE) describing 
the temperature response to the heat flow, the second two equations 
(\ref{eq_oneDrodBnhA}) and (\ref{eq_oneDrodBnhB}) specify the  
boundary conditions (BC), each of which which are taken to be a linear 
combination
of Neumann and Dirichlet boundary conditions. The final equation
(\ref{eq_oneDrodCnh}) is the initial condition (IC), specifying the temperature 
profile of the rod at $t=0$.  When the right-hand sides of the BC\rq{}s
vanish, $\gamma_1= \gamma_2=0$, the problems is called 
{\em homogeneous}, otherwise the problem is called {\em nonhomogeneous}. 
The special property of homogeneous problems is that the sum 
of any two homogeneous solutions is another homogeneous solution. 
This is not true of nonhomogeneous problems, since the nonhomogeneous
BC will not be satisfied by the sum of two nonhomogeneous solutions.

Finding a solution to the nonhomogeneous problem 
(\ref{eq_oneDrodAnh})--(\ref{eq_oneDrodCnh}) involves two steps. The first 
is to find a general solution to the homogeneous problem, which 
Wdenote by $\tilde T(x,t)$
in the text; and the second step is to find a specific solution to the
nonhomogeneous problem. We accomplish the latter by finding a {\em  static}
nonhomogeneous solution, which is denoted by $\bar T(x)$, as this is easier 
than finding a fully dynamic nonhomogeneous solution.\footnote{
\footnoteskip
This involves solving the linear equation $\partial^2 T/\partial x^2=0$ in 1D, 
and Laplace\rq{}s equation $\nabla^2 T =0$ in 2D.
}
There are times when
finding a static nonhomogeneous solution is not possible, but in our
context, these cases are rare, and will not be treated here. The sum of 
the general 
homogeneous and the specific nonhomogeneous solutions,
\begin{eqnarray}
  T(x,t) = \tilde T(x,t) + \bar T(x)
  \ ,
  \label{Tgeneral}
\end{eqnarray}
will in fact be a 
solution to the full nonhomogeneous problem. The homogeneous solution
$\tilde T(x,t)$ will be represented as a Fourier series, and its coefficients 
will be chosen so that the initial condition (\ref{eq_oneDrodCnh}) is satisfied
by $T(x,t)$, {\em i.e.} we choose the Fourier coefficients of $\tilde T$ such 
that
\begin{eqnarray}
  \tilde T(x,0) = T_0(x) - \bar T(x) 
  \ .
  \label{TBC}
\end{eqnarray}

The boundary conditions (\ref{eq_oneDrodBnhA}) and (\ref{eq_oneDrodBnhB})
are specified by the coefficients $\alpha_i$,
$\beta_i$, and $\gamma_i$ for $i=1,2$. Combinations of these parameters
produce temperatures and fluxes $T_i$ and $F_i$, and it is often more 
convenient  to specify the boundary conditions in terms of these quantities.
For example, if $\beta_1=0$ in (\ref{eq_oneDrodBnhA}), then the
BC becomes $\alpha_1 T(0,t) = \gamma_1$, which we can rewrite
in the form $T(0,t)=T_1$ with $T_1 = \gamma_1/\alpha_1$. This
leads to four special cases for the boundary condition, the first being
\begin{eqnarray}
  && {\rm BC1}
  \nonumber \\
  &&
  T(0,t) =T_1 ~:~
  \hskip0.15cm 
  \alpha_1 \ne 0 \hskip0.5cm \beta_1 = 0 \hskip0.5cm \gamma_1 \ne 0 
  \hskip1.0cm T_1 = \frac{\gamma_1}{\alpha_1}
  \label{BCTone}
  \\[5pt]
  &&
  T(L,t) = T_2 ~:~
  \hskip0.15cm 
  \alpha_2 \ne 0 \hskip0.5cm \beta_2 = 0  \hskip0.5cm \gamma_2 \ne 0 
  \hskip1.0cm  T_2 = \frac{\gamma_2}{\alpha_2}
  \label{BCTtwo}  
  \ .
\end{eqnarray}
By setting $\alpha_1=\alpha_2=0$, with $\beta_i \ne 0$, we arrive at the 
heat flux boundary condition,
\begin{eqnarray}
  &&{\rm BC2}
  \nonumber \\
  &&
  \partial_x T(0,t) = F_1 ~:~
  \hskip0.15cm
  \alpha_1 = 0 \hskip0.5cm \beta_1 \ne 0 \hskip0.5cm \gamma_1 \ne 0 
  \hskip1.0cm  F_1 = \frac{\gamma_1}{\beta_1}
  \label{eq_BC2A_non}
  \\[5pt]
  &&
  \partial_x T(L,t) = F_2 ~:~
  \hskip0.15cm
  \alpha_2 = 0 \hskip0.5cm \beta_2 \ne 0   \hskip0.5cm \gamma_2 \ne 0   
  \hskip1.0cm  F_2 = \frac{\gamma_2}{\beta_2}  
  \label{eq_BC2B_non}
  \ .
\end{eqnarray}
As we shall see, we must further constrain the heat flux so that $F_1=F_2$.
This is because in a static configuration, the heat flowing into the system
must equal the heat flowing out of the system.  Finally, we can set a 
temperature boundary condition at one end of the rod, and a flux boundary
condition at the other. This can be performed in two ways,
\begin{eqnarray}
  {\rm BC3}  \hskip1.3cm &&
  \nonumber \\
  T(0,t) = T_1 &&:
  \hskip0.15cm 
  \alpha_1 \ne 0 \hskip0.5cm \beta_1 = 0 \hskip0.5cm \gamma_1 \ne 0 
  \hskip1.0cm T_1 = \frac{\gamma_1}{\alpha_1}
  \label{BC3nonhomoA}
  \\[5pt]
  \partial_x T(L,t) = F_2 &&:
  \hskip0.15cm 
  \alpha_2 = 0 \hskip0.5cm \beta_2 \ne 0   \hskip0.5cm \gamma_2 \ne 0
  \hskip1.0cm T_2 = \frac{\gamma_2}{\alpha_2}  
  \label{BC3nonhomoB}
  \ ,
\end{eqnarray}
or
\begin{eqnarray}
  {\rm BC4} \hskip1.5cm &&
  \nonumber \\
  \partial_x T(0,t) = F_1 && : 
  \hskip0.15cm
  \alpha_1 = 0 \hskip0.5cm \beta_1 \ne 0 \hskip0.5cm \gamma_1 \ne 0 
  \hskip1.0cm   F_1 = \frac{\gamma_1}{\beta_1}
  \label{BC4nonhomoA}
  \\[5pt]
  T(L,t) =T_2 &&:
  \hskip0.15cm 
  \alpha_2 \ne 0 \hskip0.5cm \beta_2 = 0   \hskip0.5cm \gamma_2 \ne 0 
  \hskip1.0cm   T_2 = \frac{\gamma_2}{\alpha_2}  
  \label{BC4nonhomoB}  
  \ .
\end{eqnarray}
Note that BC3 and BC4 are physically equivalent, and represent a rod that has 
been flipped from left to right about its center. In the following sections, 
we shall compute the solution for each of boundary conditions 
BC1 $\cdots$ BC4, as well as the case of general BC\rq{}s.

While the heat flow problem is well defined and solvable for arbitrary 
(continuous) 
profiles $T_0(x)$, a particularly convenient choice of an initial condition 
is the linear function 
\begin{eqnarray}
  T_0(x) 
  = 
  T_0^\text{lin}(x; T_\smL, T_\smR) 
  = 
  T_\smL + \frac{T_\smR - T_\smL}{L}\, x
  \ ,
 \label{IClinear}
\end{eqnarray}
where $T_\smL$ is the initial temperature at the far left of the rod, 
$x=0^+$, and $T_\smR$ is the initial temperature at the far right
of the rod, $x=L^-$. 
We have used the notation $x=0^+$ and $x=L^-$ because the 
initial condition only holds on the open interval $0 < x < L$, and, strictly 
speaking, $T_0(x)$ is not defined at $x=0$ and $L$, as this would 
``step on\rq\rq{} the boundary conditions at these end-points (the
system would be over constrained at $x=0,L$). This 
leads to the interesting possibility that the initial condition can be 
incommensurate with the boundary conditions, in that $T_\smL$ need 
not agree with $T_1$, nor $T_\smR$ with $T_2$. 

Taking the boundary condition BC1 for definiteness, let us examine 
the resulting solution $T(x,t)$ when  $T_1 \ne T_\smL$ or 
$T_2 \ne T_\smR$. If we consider such a solution $T(x,t)$ on the 
open \hbox{$x$-interval} $(0,L)$, then $T(x,t)$ converges to the initial 
profile $T_0(x)$ as $t$ goes to zero, that is to say, $T(x,t) \to T_0(x)$ 
as $t \to 0$ for all $x \in (0,L)$; however, this point-wise convergence 
is {\em nonuniform}.  See Ref.~\cite{Rubin1976} for an introductory 
but solid treatment of real analysis and uniform convergence, and 
Appendix B for a short summary of uniform convergence. Alternatively, 
we may consider the solution $T(x,t)$ on the closed interval $[0,L]$ by
appending the boundary conditions at $x=0,L$. Then the limit of $T(x,t)$ 
as $t \to 0$ is a the function taking the values $T=T_1$ at $x=0$, 
$T=T_2$ at $x=L$, and $T=T_0(x)$  at $x \in (0,L)$. If $T_1 \ne T_\smL$ 
or $T_2 \ne T_\smR$, the limit function $\lim_{t \to 0} T(x,t)$ is 
discontinuous at $x=0,L$, even though every function $T(x,t)$ in the 
sequence is continuous in $x$. We have therefore found a sequence of 
{\em continuous} functions $T(x,t)$ (continuous in $x$ and indexed by 
$t$) whose limit is a {\em discontinuous} function, and this is exactly 
what one would expect of a nonuniformly converging sequence of 
functions. Not surprisingly, if we set the boundary condition to agree 
with the initial condition, $T_1 =T_\smL$ and $T_2 = T_\smR$, then 
the limit function is continuous; however, the initial condition $T_0(x)$ 
becomes a static nonhomogeneous solution to the heat equations.  
\subsection{Some Heat Flow Problems in ExactPack}

\begin{figure}[t!]
\includegraphics[scale=0.39]{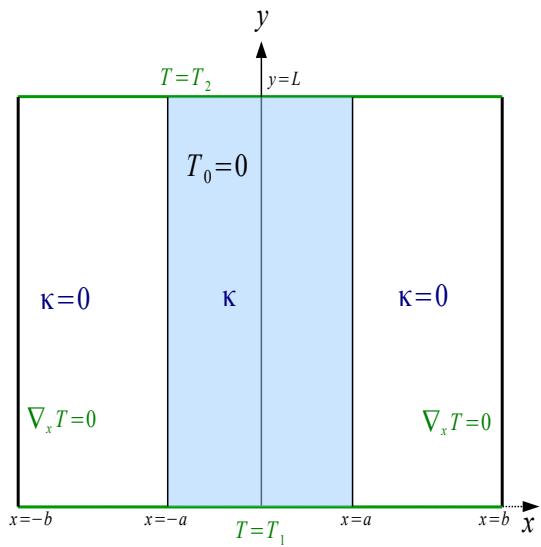} 
\vskip-2.5cm
\caption{\footnoteskip  
The Planar Sandwich. The inner material in blue (the meat) located within 
$-a \le x \le a$ is heat conducting with $\kappa > 0$. The outer 
materials (the bread), located within $-b \le x < -a$ and $a < x \le b$, 
are not heat conducting and have $\kappa=0$. The boundary temperature 
is uniform in $x$ along the lower and upper boundaries, with temperatures
$T(x,0)=T_1$ and $T(x,L)=T_2$. The temperature flux along the far left and 
right boundaries vanishes, $\partial_x T(\pm b,y)=0$. Finally, the initial
temperature is taken to be $T_0(x,y)=0$ inside the entire region 
$(-b,b)\times (0,L)$.
}
\label{fig_planar_sandwich}
\end{figure}

\begin{figure}[t!]
\includegraphics[scale=0.40]{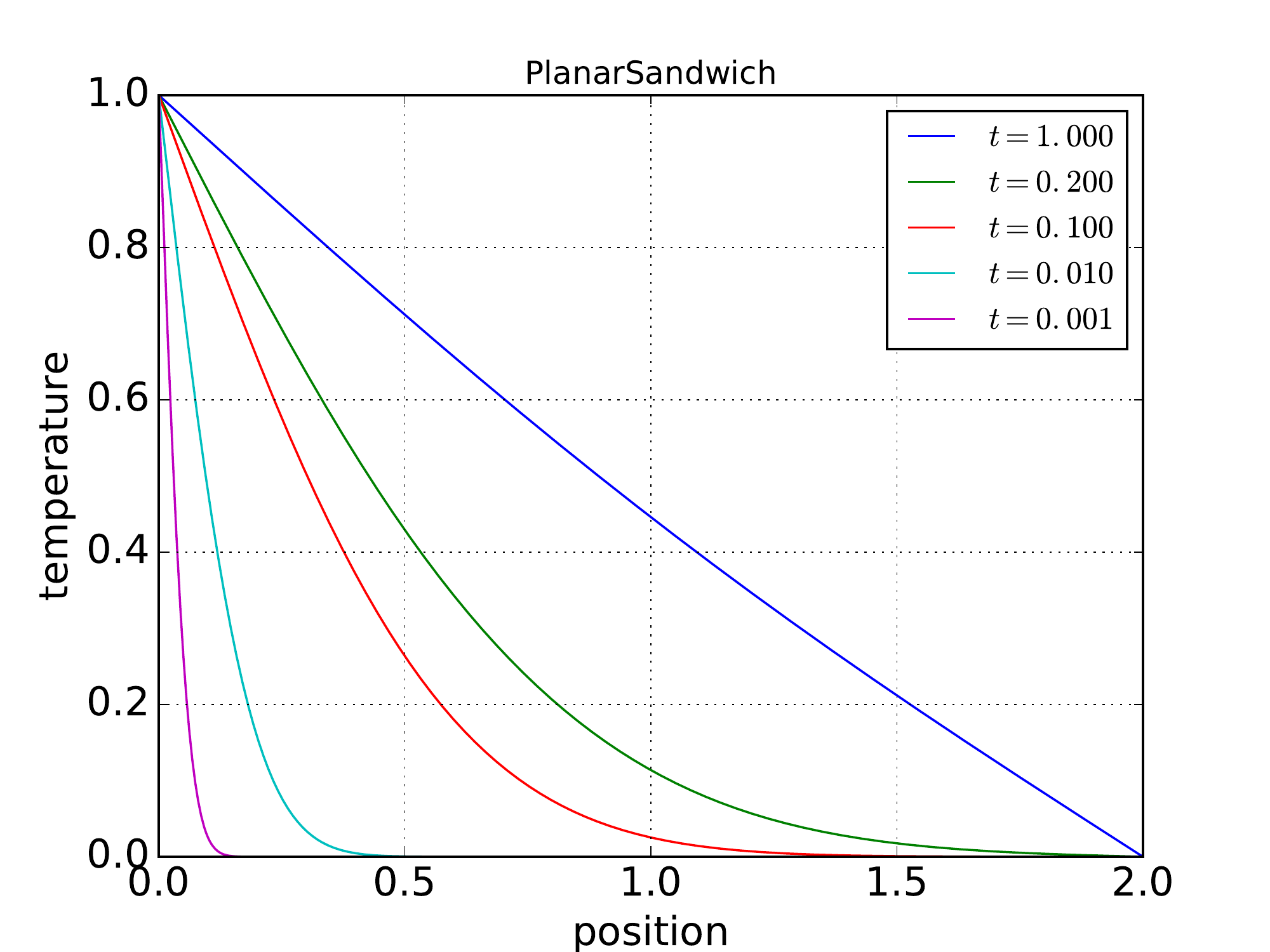} 
\caption{\footnoteskip  
The Planar Sandwich in ExactPack: PlanarSandwich(T1=1, T2=0, L=2, 
Nsum=1000). The temperature profile is plotted at times
$t=1, 0.2, 0.1, 0.01$, and $0.001$. The BC\rq{}s are $T(0)=1$, T(L)=0, 
and The IC is $T_0=0$. The diffusion constant is $\kappa=1$, the length 
of the rod is $L=2$, and we have summed over 1000 terms in the series
}
\label{fig_planar_sandwich_ep}
\end{figure}

The first test problem of Ref.~\cite{Dawes2016} is a heat flow problem 
in 2D rectangular coordinates called the Planar Sandwich, illustrated
in Fig.~\ref{fig_planar_sandwich}. The problem consists of three material 
layers aligned along the y-direction in a sandwich-like configuration. 
The outer two layers do not conduct heat ($\kappa=0$), while the inner 
layer is heat conducting with $\kappa>0$, forming a sandwich of 
conducting and non-conducting materials. The temperature boundary 
condition on the lower $y=0$ boundary is taken to be $T(x,y\!=\!0)=T_1$, 
while the temperature on the upper boundary is $T(x,y\!=\!L)=T_2$. The 
temperature flux in the $x$-direction on the far left and right ends of 
the sandwich vanishes, $\partial_x T(\pm b, y)=0$. Finally, the initial
temperature inside the sandwich is taken to vanish, $T_0(x,y)=0$. 
Symmetry arguments reduce the problem
to 1D heat flow in the $y$-direction, and in this subsection we shall 
orient the 1D rod of the previous section along the $y$-direction  
rather than the $x$-direction (in the remaining sections, however,
we shall revert to the convention of heat flow along $x$). This brief 
change in convention allows us to keep with the original notation 
defined in Ref.~\cite{Dawes2016}. The heat flow equation in the
central region, $\vert x \vert \le a$, reduces to 1D flow
along the $y$-direction,
\begin{eqnarray}
  \frac{\partial T}{\partial t}
  &=&
  \kappa\, \frac{\partial^2 T}{\partial y^2}
  \ .
\end{eqnarray}
We now represent the temperature profile as a function of $y$, so that
$T= T(y,t)$, and the boundary conditions of the rod become $T(0,t)=T_1$ 
and $T(L,t)=T_2$,  as in BC1. The initial condition becomes $T_0(y)=0$. 
The exact analytic solution was presented in Ref.~\cite{Dawes2016}, and 
takes the form
\begin{eqnarray}
  T(y,t)
  &=&
  T_1 + \frac{(T_2 - T_1) \, y}{L}
  + 
  \sum_{n=1}^\infty B_n \, \sin(k_n y) \, e^{-\kappa\, k_n^2 t}
  \\[5pt]
  k_n &=& \frac{n \pi}{L} 
  \hskip0.5cm {\rm and}\hskip0.5cm
  B_n 
  =
  \frac{2 T_2 (-1)^n - 2 T_1}{n\pi}
  \ ,
  \label{eq_planar_sandwich}
\end{eqnarray}
for $\vert x \vert \le a$; and $T=0$ for $\vert x \vert > a$. 
Figure~\ref{fig_planar_sandwich_ep} illustrates a plot of the planar 
sandwich solution for the initial conditions $T_1=1$ and $T_2=0$, 
at several representative times \hbox{$t=1 \,,\, 0.2 \,,\,  0.1 \,,\, 0.01,$}
and $0.001$. The instance of the planar sandwich 
class used to plot the figure was created by the python call

\vskip0.2cm
\noindent
\verb+solver = PlanarSandwich(T1=1, T2=0, L=2, Nsum=1000) + \ .
\vskip0.2cm

\noindent
This class instance sets the boundary conditions to $T_1=1$ and $T_2=0$, 
the length of the rod to $L=2$, and it sums over the first 1000 terms of the 
series. By default it also sets the IC to $T_0=0$. For each of the five 
representative values of $t$, we must create five solution objects,~{\em i.e.}

\vskip0.2cm
\footnoteskip
\noindent
\vbox{
\begin{verbatim}
t0 = 0.001
t1 = 0.01
...
soln0 = solver(y, t0)
soln1 = solver(y, t1)
...  ,
\end{verbatim}
}
\bodyskip

\vskip-0.25cm
\noindent
where \verb+y+ is an array of grid values ranging from $y=0$ to $y=L=2$. 
The solutions can then be plotted in the standard ExactPack manner, 
\verb+soln0.plot()+, \verb+soln1.plot()+, {\em etc.} The script that 
produces the plot in Fig.~\ref{fig_planar_sandwich_ep} is given in 
Appendix~\ref{sec:sample_ep_script}. 

\begin{figure}[t!]
\includegraphics[scale=0.40]{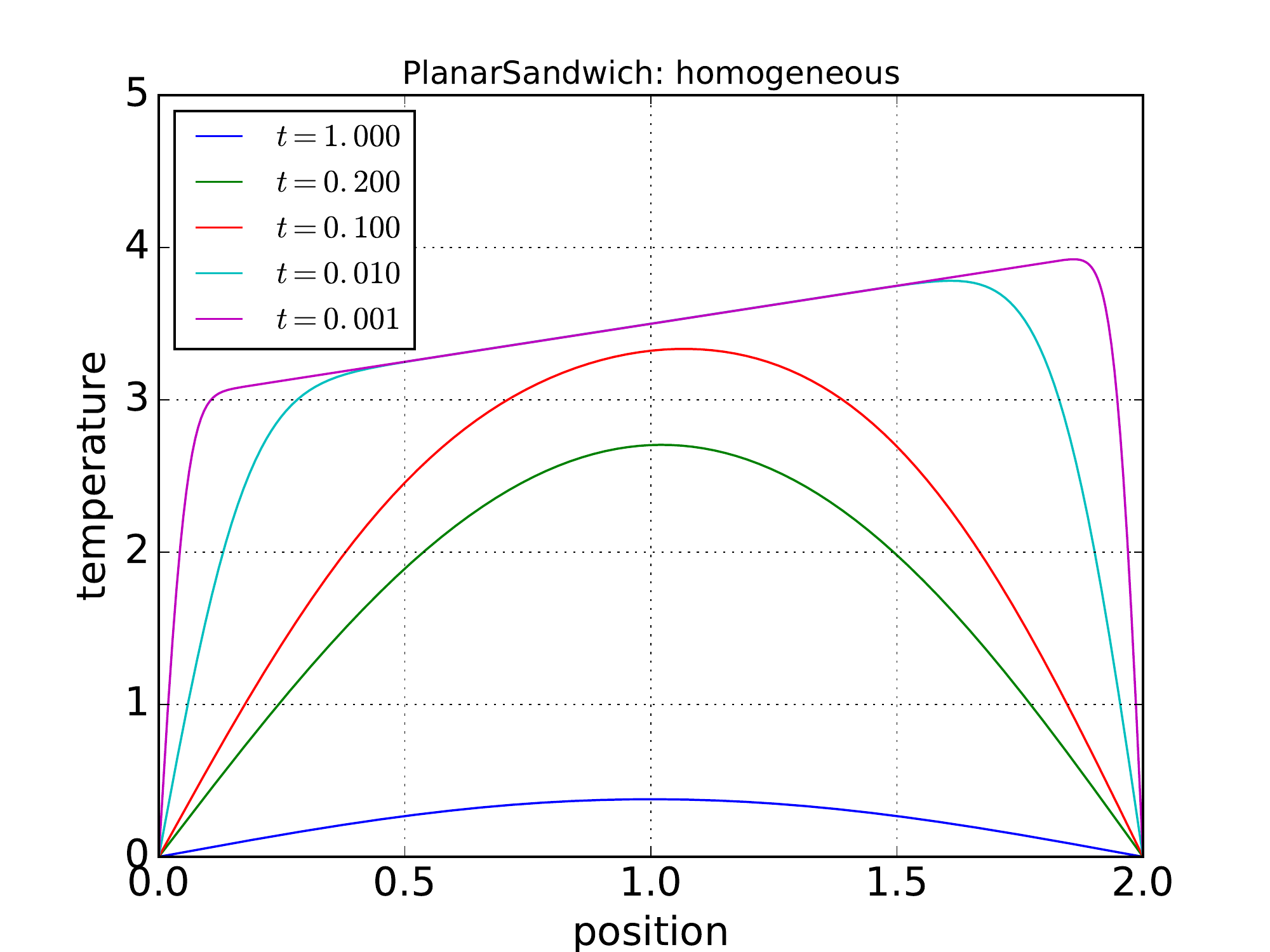} 
\caption{\footnoteskip  
The Planar Sandwich: PlanarSandwich(T1=0, T2=0, TL=3, TR=4, L=2, 
Nsum=1000). 
Temperature profiles for the homogeneous planar sandwich at times 
$t=1, 0.2, 0.1, 0.01$, and $0.001$,  with $\kappa=1$, $L=2$, 
$T_\smL=3$, $T_\smR=4$ (and $T_1=T_2=0$).  The boundary conditions $T_1=0$ and $T_2=0$ render 
the solution homogenous, while the initial condition $T_0(y)$, specified 
by $T_\smL$ and $T_\smR$, specifies the linear function (\ref{ICliny})
as the initial condition. As $t \to 0$, the solution $T(y,t)$ convergens 
nonuniformly on the open $y$-interval $(0,L)$ to $T_0(y)$.
}
\label{fig_planar_sandwich_homo_ep}
\end{figure}

In the following sections, we shall analyze heat flow in a 1D rod in some 
detail, and we will see that  by modifying the boundary conditions, as well 
as the initial condition, we can form a number of variants of the planar 
sandwich. In our first variant, we take $T_1=0$ and $T_2=0$ (the 
homogeneous version of BC1), but 
we choose a nontrivial initial condition for $T_0(y)$. An arbitrary continuous 
function would suffice, but for simplicity we employ a linear 
initial condition for $T_0(y)$. Since, in this section, the heat flow is along 
the $y$-direction, the linear initial condition (\ref{IClinear}) must be 
translated into
\begin{eqnarray}
  T_0(y) = T_0^\text{lin}(y) = T_\smL + \frac{T_\smR - T_\smL}{L}\, y
  \ .
  \label{ICliny}
\end{eqnarray}
As shown in the next section, the solution takes the form
\begin{eqnarray}
 T(y,t)
 &=&
 \sum_{n=1}^\infty 
 B_n \, \sin(k_n y)\, e^{-\kappa \, k_n^2 t}
 \\[5pt]
  k_n &=& 
  \frac{n \pi}{L}
  \hskip0.5cm {\rm with}\hskip0.5cm
  B_n 
  =
  \frac{2T_\smL - T_\smR (-1)^n}{n\pi}
  \ .
  \label{eq_variant_two}
\end{eqnarray}
This is illustrated in Fig.~\ref{fig_planar_sandwich_homo_ep} 
for the initial condition specified by $T_\smL=3$ and $T_\smR=4$.
For this case, the class PlanarSandwich is instantiated by

\vskip0.2cm
\noindent
 \verb+solver = PlanarSandwich(T1=0, T2=0, TL=3, TR=4, L=2, Nsum=1000)+ \ .
\vskip0.2cm

\noindent
The similarity between the coefficients $B_n$ in (\ref{eq_variant_two}) and
(\ref{eq_planar_sandwich}) is somewhat accidental, and arises from the 
choice of the linear initial condition (\ref{ICliny}), which, coincidentally, 
is the same form as the nonhomogeneous solution $\bar T(x)$ used to 
construct the original variant of the planar sandwich (\ref{eq_planar_sandwich}). 
It is this that accounts 
for the similarity. This example also illustrates how to override the default
parameters in an ExactPack class, in this case, by setting $T_1=0$ and 
$T_2=0$.  The default initial condition is $T_0(y)=0$,  and this is why we 
did not need to specify the values of $T_\smL$ and $T_\smR$ in
Fig.~\ref{fig_planar_sandwich_ep}, and  why we had to override
these values in Fig.~\ref{fig_planar_sandwich_homo_ep}.

As another variant on the planar sandwich, we can choose vanishing heat flux 
on the upper and lower boundaries (as in BC2). This will be called the Hot 
Planar Sandwich, in analogy with the Hot Cylindrical Sandwich of 
Ref.~\cite{Dawes2016}, and its solution takes the form
\begin{eqnarray}
 T(y,t)
 &=&
 \frac{A_0}{2} + \sum_{n=1}^\infty 
 A_n \, \cos(k_n y)\, e^{-\kappa \, k_n^2 t}
 \\[5pt]
  k_n &=& 
  \frac{n \pi}{L}
  \\[5pt]
  A_0 &=& 
  \frac{T_\smL + T_\smR}{2}
  \hskip0.5cm {\rm and~for}~ n \ne 0,\hskip0.5cm
  A_n 
  =
  2 \Big(T_\smL - T_\smR \Big)\frac{1 - (-1)^n}{n^2\pi^2}
  \ .
\end{eqnarray}
This new variant of the planar sandwich can be instantiated by

\vskip0.2cm
\noindent
\verb+solver = PlanarSandwichHot(F=0, TL=3, TR=3, L=2, Nsum=1000)+ \ .
\vskip0.2cm

\begin{figure}[h!]
\includegraphics[scale=0.40]{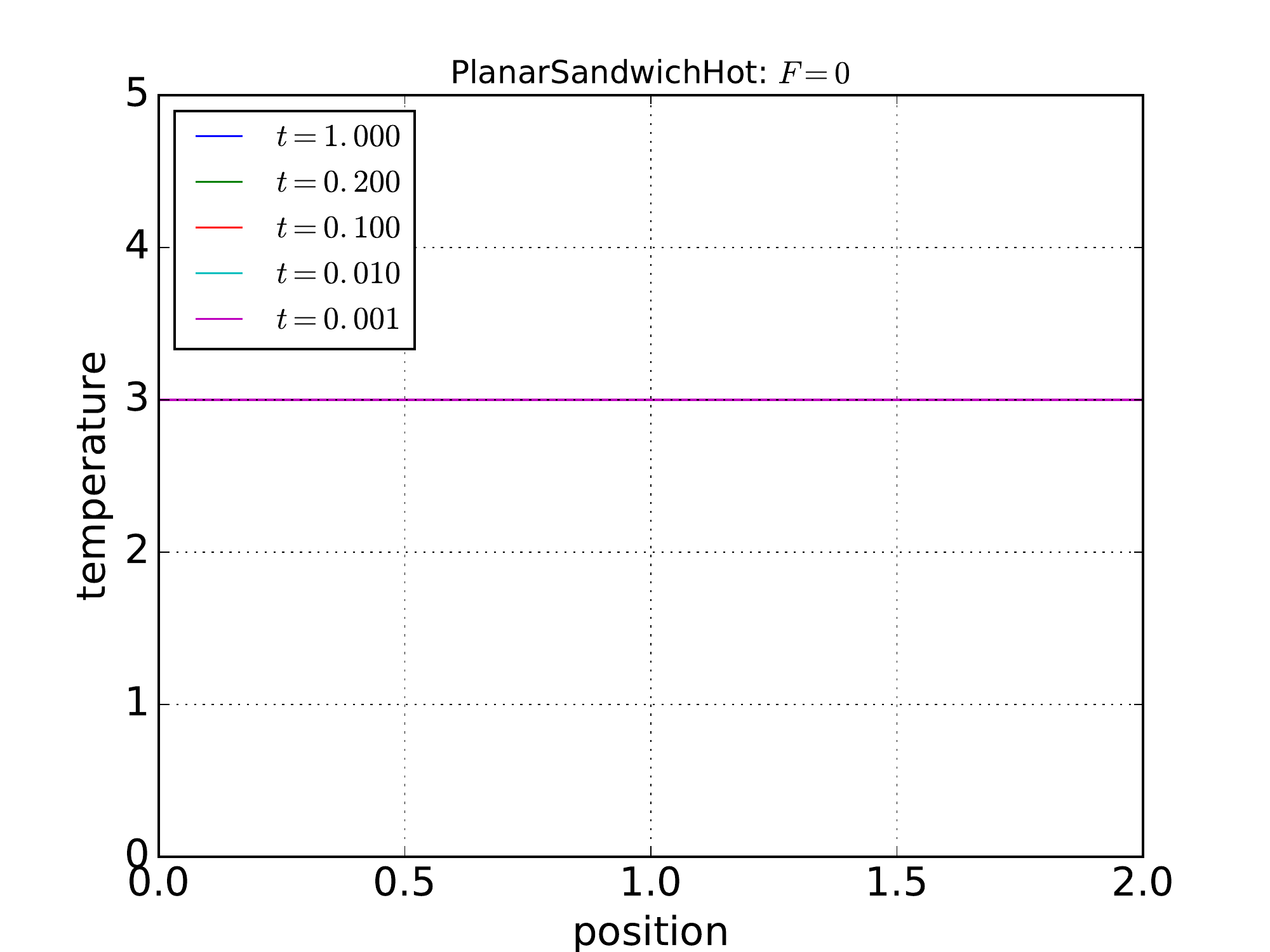} 
\caption{\footnoteskip  
The Hot Planar Sandwich in ExactPack: PlanarSandwichHot(F=0, TL=3, 
TR=3, L=2, Nsum=1000). Since the heat flux on the boundaries vanishes, 
heat cannot escape from the material, and the temperature must remain
constant in time. The temperature profile has been plotted for the 
times $t=1, 0.2, 0.1, 0.01$, and $0.001$, and is indeed constant.
}
\label{fig_planar_sandwich_hot_ep}
\end{figure}

\begin{figure}[h!]
\includegraphics[scale=0.40]{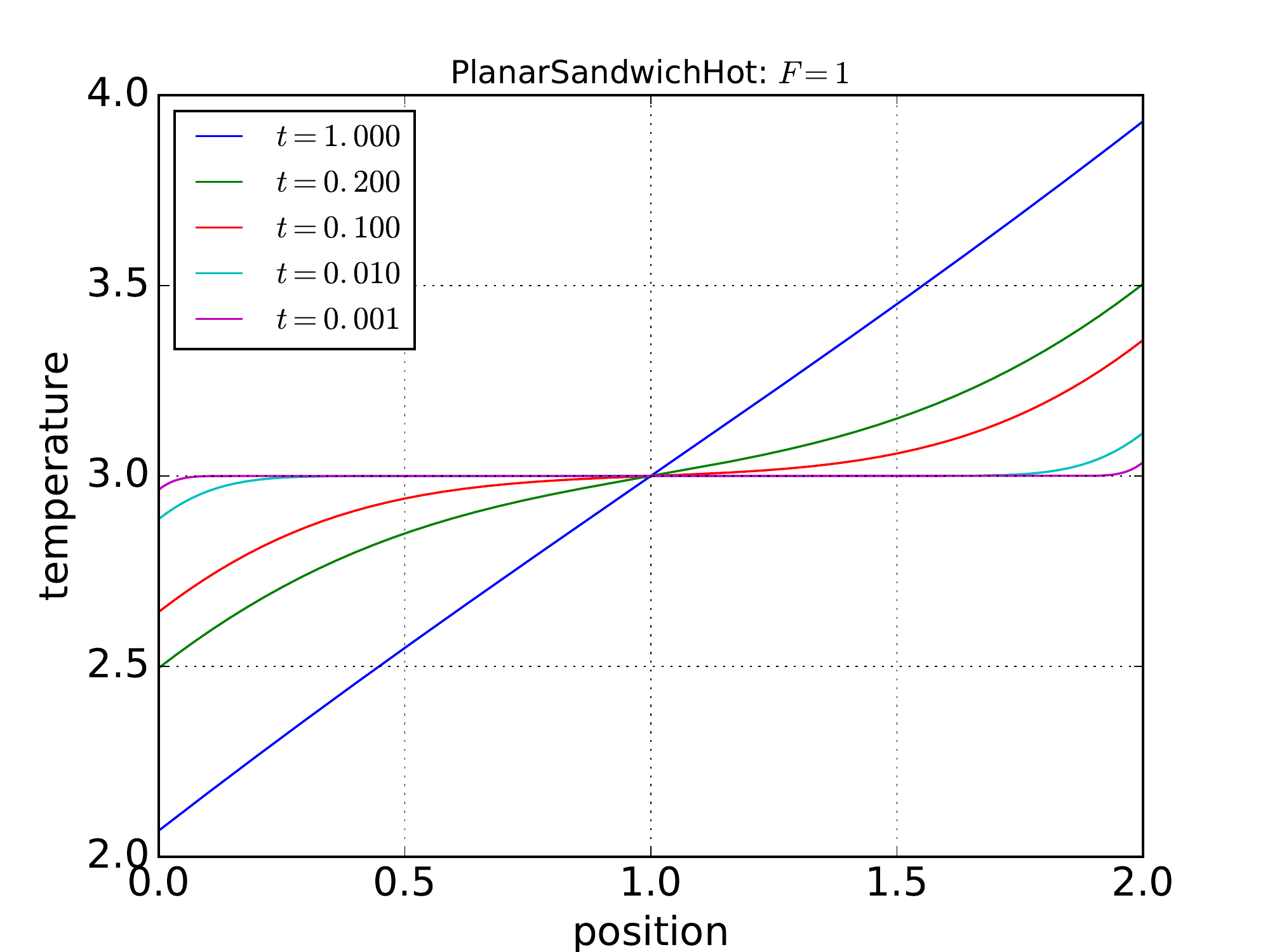} 
\caption{\footnoteskip  
The Hot Planar Sandwich in ExactPack: PlanarSandwichHot(F=1, TL=3, 
TR=3, L=2, Nsum=1000). The profiles are plotted for times
$t=1, 0.2, 0.1, 0.01$, and $0.001$. The heat flux at the boundaries 
is $F=1$, and we see that the temperature profile changes as heat
flows out of the rod.
}
\label{fig_planar_sandwich_ho_2t_ep}
\end{figure}

\noindent
The heat flux $F$ on the boundaries has been set to zero, and a constant initial 
condition $T_0=3$ has been specified (by setting  $T_\smL=T_\smR=3)$.
The solution is illustrated in Fig.~\ref{fig_planar_sandwich_hot_ep}. 
On physical grounds, heat cannot escape from the material, 
and the temperature  must remain constant.  In contrast, when the
heat flux is nonzero, heat is free to flow from the sandwich to the 
environment, and the temperature need not remain constant. For 
a flux $F=1$, the change in the temperature profiles with time is 
illustrated in  Fig.~\ref{fig_planar_sandwich_ho_2t_ep}. 

\begin{figure}[h!]
\includegraphics[scale=0.40]{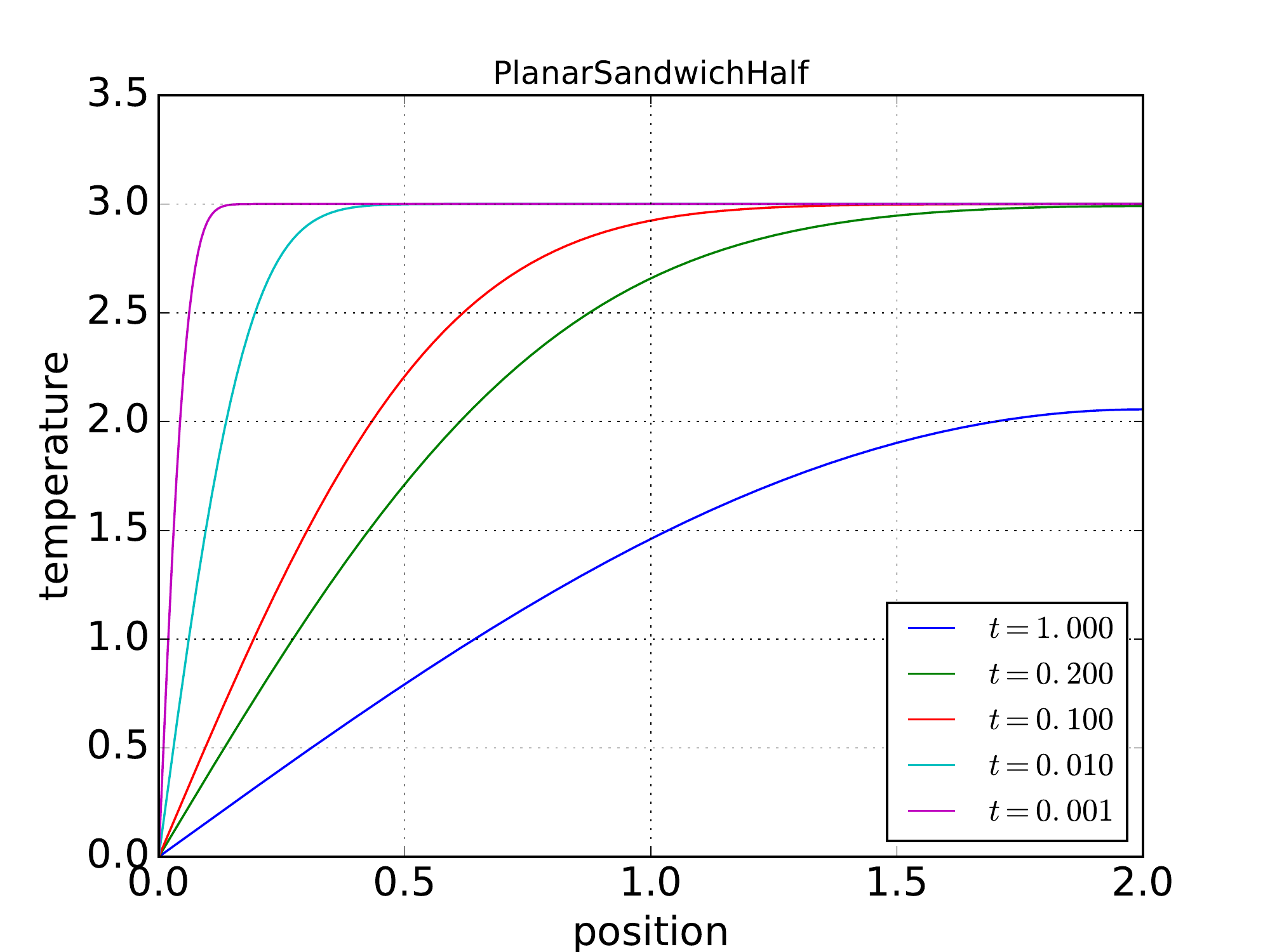} 
\caption{\footnoteskip  
The Half Planar Sandwich in ExactPack: PlanarSandwichHalf(T=0, F=0, 
TL=3, TR=3, L=2, Nsum=1000). The profiles are plotted for times
$t=1, 0.2, 0.1, 0.01$, and $0.001$. Note that the profiles clearly
satisfy the temperature on the left vanishes, and the derivative
of the temperature on the right vanishes.
}
\label{fig_planar_sandwich_half_ep}
\end{figure}

Another variant on the planar sandwich is to choose vanishing heat flux
on the upper boundary,  $\partial_y T(L)=0$, and zero temperature on
the lower boundary, $T(0)=0$. This is an example of boundary condition
BC3, and the solution is called the Half Planar Sandwich. As we show in 
the next section, the solution takes the form
\begin{eqnarray}
 T(y,t)
 &=&
 \sum_{n=0}^\infty 
 B_n \, \sin(k_n y)\, e^{-\kappa \, k_n^2 t}
 \\[5pt]
  k_n &=& 
  \frac{(2 n + 1) \pi}{L}
  \hskip0.5cm {\rm with}\hskip0.5cm
  B_n 
  =
  \frac{4 T_\smR}{(2 n + 1) \pi}
  -
  \frac{8\big(T_\smR - T_\smL\big) }{(2n+1)^2 \pi^2} 
  \ .
\end{eqnarray}
Taking the initial condition $T_0=3$ ($T_\smL=T_\smR=3)$ gives
Fig.~\ref{fig_planar_sandwich_half_ep}, which is instantiated by
\vskip0.2cm
\noindent
\verb+solver = PlanarSandwichHalf(T=0, F=0, TL=3, TR=3, L=2, Nsum=1000)+ \ .
\vskip0.2cm
\noindent
If we had chosen $\partial_y T(0)=0$ and $T(L)=0$, as in BC4, then 
the figure would have been reflected about the central point $y=1$,
but otherwise physically identical.

\pagebreak
\section{The Static Nonhomogeneous Problem}
\label{sec:nonhomogeneous}

As previously discussed, the full nonhomogeneous problem is divided 
into two parts: (i)~finding a general homogeneous solution $\tilde T(x,t)$,
and (ii) finding a specific nonhomogeneous static solution $\bar T(x)$. 
Because of
its simplicity, we first turn to solving the corresponding  nonhomogeneous
equations. We start with the {\em static} or equilibrium heat equation for 
$\bar T(x)$ with nonhomogeneous BC\rq{}s, 
\begin{eqnarray}
  {\rm DE}: 
  \hskip2.9cm
  \frac{\partial^2 \bar T(x)}{\partial x^2}
  &=&
  0
  \hskip3.0cm 
  0 < x < L 
\label{eq_oneDrodALP}
\\[5pt]
  {\rm BC}:  
  \hskip1.16cm
  \alpha_1 \bar T(0) + \beta_1 \bar T^\prime(0) &=& \gamma_1
\label{eq_oneDrodBLP}  
\\[-3pt]
  \alpha_2 \bar T(L) + \beta_2 \bar T^\prime(L) &=& \gamma_2
  \ .
\label{eq_oneDrodCLP}
\end{eqnarray}
The solution to (\ref{eq_oneDrodALP}) is trivial, and may be
written in the form,
\begin{eqnarray}
  \bar T(x) = a + b \, x
  \ ,
  \label{eq_barTab}
\end{eqnarray}
or alternatively,
\begin{eqnarray}
  \bar T(x) &=& T_1 + \frac{T_2 - T_1}{L}\, x
  \ .
 \label{eq_barTlr}
 \end{eqnarray}
The coefficients $a$ and $b$, or $T_1$ and $T_2$, are determined 
by the nonhomogeneous boundary conditions (\ref{eq_oneDrodBLP})  
and (\ref{eq_oneDrodCLP}). Note that, coincidentally, that the static
nonhomogeneous solution $\bar T(x)$ takes the same form as the 
linearized initial condition of (\ref{IClinear}),
namely,
\begin{eqnarray}
  \bar T(x) &=& T_0^\text{lin}(x; T_1, T_2)
  \ .
  \label{barTT0lin}
 \end{eqnarray}
While this is a fortuitous coincidence of 1D heat flow, and does not hold 
for 2D heat 
flow, (\ref{barTT0lin}) will be used in the following sections to simplify 
the algebra in calculating expansion coefficients for the homogenous 
and nonhomogeneous solutions. We turn now to finding the appropriate 
values of $T_1$ and $T_2$ for the case of general boundary conditions, 
and then for the four special cases,

\vskip0.4cm 
\begin{enumerate}
  \baselineskip 10pt plus 1pt minus 1pt
  \setlength{\itemsep}{3pt} 
  \setlength{\parskip}{1pt} %
  \setlength{\parsep}{0pt}  %
\item[BC1:] (\ref{BCTone})--(\ref{BCTtwo})
\item[BC2:] (\ref{eq_BC2A_non})--(\ref{eq_BC2B_non})
\item[BC3:] (\ref{BC3nonhomoA})--(\ref{BC3nonhomoB})
\item[BC4:]  (\ref{BC4nonhomoA})-(\ref{BC4nonhomoB}) \ .
\end{enumerate}
\subsection{General Boundary Conditions}

As exhibited in  (\ref{eq_barTab})--(\ref{eq_barTlr}), the
nonhomogeneous solution $\bar T(x)$ can be expressed in the form
\begin{eqnarray}
  \bar T(x) 
  = 
  a + b\, x
  =
  T_1 + \frac{T_2 - T_1}{L}\, x
  \ ,
\end{eqnarray}
where $\bar T(0)=a= T_1$ and $\bar T(L) = a + b L = T_2$. The BC\rq{}s
(\ref{eq_oneDrodBLP}) and (\ref{eq_oneDrodCLP}), and the solution
 (\ref{eq_barTab}), reduce to a linear equation in terms of $a$ and $b$,
\begin{eqnarray}
   \left(
   \begin{array}{cc}
   \alpha_1  & \beta_1 \\
   \alpha_2  &  ~\beta_2 + \alpha_2L 
  \end{array}
  \right)
  \left(
  \begin{array}{c}
  a \\
  b
  \end{array}
  \right)
  =
    \left(
  \begin{array}{c}
  \gamma_1 \\
  \gamma_2
  \end{array}
  \right)
  \ .
\end{eqnarray}
Upon solving this equation we find
\begin{eqnarray}
  a 
 &=& 
 \frac{-\beta_1 \gamma_2 + \beta_2 \gamma_1 + L \alpha_2 \gamma_1}
 {\alpha_1 \beta_2 - \alpha_2 \beta_1 + L \alpha_1 \alpha_2 }
  \\[5pt] 
  b 
  &=&  
  \frac{\alpha_1 \gamma_2 - \alpha_2 \gamma_1}
  {\alpha_1 \beta_2 - \alpha_2 \beta_1 + L \alpha_1 \alpha_2}
  \ ,
\end{eqnarray}
or in terms of temperature parameters, $T_1=a$ and $T_2=a + b L$,
we can write
\begin{eqnarray}
  T_1
 &=& 
 \frac{\beta_2 \gamma_1 -\beta_1 \gamma_2 + L \alpha_2 \gamma_1}
 {\alpha_1 \beta_2 - \alpha_2 \beta_1 + L \alpha_1 \alpha_2 }
 \label{T1genBC}
  \\[5pt] 
  T_2
  &=&  
 \frac{\beta_2 \gamma_1 -\beta_1 \gamma_2 + L \alpha_1 \gamma_2}
 {\alpha_1 \beta_2 - \alpha_2 \beta_1 + L \alpha_1 \alpha_2 }
  \label{T2genBC}
\ .
\end{eqnarray}
Note that the determinant of the linear equations vanishes for BC2, 
and we must handle this case separately.

\subsection{Special Cases of the Static Problem}

\subsubsection{BC1}

The first special boundary condition is (\ref{BCTone}) and (\ref{BCTtwo}),
\begin{eqnarray}
  \bar T(0) &=& T_1
  \label{bc_Tone}
  \\
  \bar T(L) &=& T_2
  \label{bc_Ttwo}
  \ ,
\end{eqnarray}
with the solution taking the form (\ref{eq_barTlr}),
\begin{eqnarray}
  \bar T(x) &=& T_1 + \frac{T_2 - T_1}{L}\, x
  \ .
 \end{eqnarray}
The temperature coefficients $T_1$ and $T_2$ are given by the temperatures 
of the upper and lower boundaries in (\ref{bc_Tone}) and (\ref{bc_Ttwo}).
Equivalently, the coefficients in (\ref{eq_barTab}) are just $a=T_1$ and 
$b = (T_2 - T_1)/L$.

\subsubsection{BC2}

Let us now find the nonhomogeneous equilibrium solution for
the boundary conditions (\ref{eq_BC2A_non}) and (\ref{eq_BC2B_non}),
\begin{eqnarray}
  \partial_x \bar T(0) &=& F_1
  \\
  \partial_x \bar T(L) &=& F_2
  \ ,
\end{eqnarray}
where $F_1$ and $F_2$ are the heat fluxes at $x=0$ and $x=L$, respectively,
and are related to the boundary condition parameters in (\ref{eq_oneDrodBLP})  
and (\ref{eq_oneDrodCLP}) by $F_1 = \gamma_1/\beta_1$ and $F_2 = 
\gamma_2/\beta_2$. As before, the general solution is $\bar T(x)=a + b x$,
and we see that $\bar T^\prime(x)=b$ is independent of $x$.  In other words, 
the heat flux at either end of the rod must be identical, $F_1 = b = F_2$. 
In fact, this result follows from energy conservation, since, in equilibrium, 
the heat flowing into the rod must be equal the heat flowing out of the rod.
Therefore, more correctly, we should have started with the boundary conditions
\begin{eqnarray}
  \partial_x \bar T(0) &=& F
  \\
  \partial_x \bar T(L) &=& F
  \label{BCFsame}
  \ ,
\end{eqnarray}
with
\begin{eqnarray}
  F = \frac{\gamma_1}{\beta_1} = \frac{\gamma_2}{\beta_2}
  \ .
\end{eqnarray}
As we saw in the previous section on general initial conditions, this 
case is singled out for special treatment.  The value of the constant term 
$a$ is not uniquely determined in this case; however, we are free to set it to 
zero, giving
\begin{eqnarray}
  \bar T(x) = F  x
  \label{barTFsame}
  \ .
\end{eqnarray}
There is nothing wrong with setting $a=0$, since we only need to find
{\em one} nonhomogeneous solution, and (\ref{barTFsame}) fits
the bill. We can write this solution in the form (\ref{eq_barTlr}), with
\begin{eqnarray}
  T_1 &=& 0
  \\
  T_2 &=&  F L 
  \ . 
\end{eqnarray}

\subsubsection{BC3}

The next set of boundary conditions are (\ref{BC3nonhomoA}) and
(\ref{BC3nonhomoB}),
\begin{eqnarray}
  \bar T(0) &=& T_1
  \\
  \partial_x \bar T(L) &=& F_2
  \ ,
\end{eqnarray}
and we can express the solution (\ref{eq_barTlr}) in terms of the temperature 
$T_1$, and the effective temperature
\begin{eqnarray}
  T_2 &=& T_1 + F_2 L
  =
  \frac{\gamma_1}{\alpha_1} + \frac{\gamma_2 L}{\beta_2}
  \ . 
\end{eqnarray}

\subsubsection{BC4}

The boundary conditions are (\ref{BC4nonhomoA}) and (\ref{BC4nonhomoB}),
\begin{eqnarray}
  \partial_x \bar T(0) &=& F_1
  \\
  \bar T(L) &=& T_2
  \ ,
\end{eqnarray}
and the solution (\ref{eq_barTlr}) can be written in terms of $T_2$ and the 
effective temperature 
\begin{eqnarray}
  T_1 &=& T_2 - F_1 L
  =
  \frac{\gamma_2}{\alpha_2} - \frac{\gamma_1 L}{\beta_1}
  \  .
\end{eqnarray}
We have now found the static homogeneous solution in the form
\begin{eqnarray}
  \bar T(x) = T_\sm1 + \frac{(T_1 - T_1)\, x}{L}
  \label{barTxFinal}
  \ ,
\end{eqnarray}
where the temperatures in  (\ref{barTxFinal}) are given by
\vskip0.3cm 
\begin{enumerate}
  \baselineskip 10pt plus 1pt minus 1pt
  \setlength{\itemsep}{3pt} 
  \setlength{\parskip}{1pt} %
  \setlength{\parsep}{0pt}  %
\item[BC1:] $T_1$ and $T_2$
\item[BC2:] $T_1=0$ and $T_2 = F L$
\item[BC3:] $T_1$ and $T_2=T_1 + F_2 l$
\item[BC4:] $T_1 = T_2 - F_1 L$ and $T_2$ \ ,
\end{enumerate}
and by (\ref{T1genBC}) and (\ref{T2genBC}) for general BC\rq{}s.

\pagebreak
\section{The Homogeneous Problem}

Now that we have found the appropriate nonhomogeneous solutions 
$\bar T(x)$, we turn to the more complicated task of finding the 
general homogeneous solutions $\tilde T(x,t)$. These solutions involve 
a Fourier sum over a discrete number of normal modes, the coefficients 
being determined by the initial conditions. These solutions depend upon 
The homogeneous equations of motion, for which $\gamma_1=0$ and
$\gamma_2=0$ in the equations 
(\ref{eq_oneDrodAnh})--(\ref{eq_oneDrodCnh}),  take the form
\begin{eqnarray}
  {\rm DE}: 
  \hskip3.73cm
  \frac{\partial \tilde T(x,t)}{\partial t}
  &=&
  \kappa\, \frac{\partial^2 \tilde T(x,t)}{\partial x^2}
  \hskip1.2cm 
  0 < x < L ~{\rm and}~ t > 0
\label{eq_oneDrodA}
\\[5pt]
  {\rm BC}:  
  \hskip1.16cm
  \alpha_1 \tilde T(0,t) + \beta_1 \partial_x \tilde T(0,t) &=& 0
  \hskip2.95cm t > 0
\label{eq_oneDrodB}  
\\[-3pt]
  \alpha_2 \tilde T(L,t) + \beta_2 \partial_x \tilde T(L,t) &=& 0
\nonumber
\\[5pt]
  {\rm IC}:  
  \hskip4.05cm
  \tilde T(x,0) &=& T_0(x)   
  \hskip2.2cm 
  0 < x < L
  \ .
\label{eq_oneDrodC}
\end{eqnarray}
As we have discussed in Section~\ref{sec:generalheatflow}, in
all of our examples we shall employ the linear initial condition
\begin{eqnarray}
  T_0(x) = T_0^\text{lin}(x; T_\smL, T_\smR) 
  = T_\smL + \frac{T_\smR - T_\smL}{L}\,x
 \label{IClinearAgain}
  \ .
\end{eqnarray}
The solution technique is by separation of variables, for which we 
assume the trial solution to be the product of independent functions 
of $x$ and $t$, 
\begin{eqnarray}
  \tilde T(x,t) = X(x) \, U(t) \ .
\end{eqnarray}
Substituting this {\em Ansatz} into the heat equation gives
\begin{eqnarray}
  \frac{dU(t)}{dt}\, X(x) = \kappa\, U(t) \,\frac{d^2X(x)}{dx^2}
  \ ,
\end{eqnarray}
or
\begin{eqnarray}
  \frac{1}{\kappa}\,\frac{U^\prime(t)}{U(t)}
  =
  \frac{X^{\prime\prime}(x)}{X(x)} 
  =
  {\rm const} 
  \equiv 
  -k^2
  \ ,
\end{eqnarray}
where we have chosen the constant to have a negative value 
$-k^2$, and we have expressed derivatives of $U(t)$ and $X(x)$ 
by primes. As usual in the separation of variables technique,
when two functions of different variables are equated, they
must be equal to a constant, independent of the variables.
The equation for $U(t)$ has the solution, 
\begin{eqnarray}
  U_k(t) &=& U_0 \, e^{-\kappa \, k^2 t}
  \ ,
\end{eqnarray}
where we have introduced a $k$-subscript to indicate that the 
solution depends upon the value of $k$. The equations for $X$  
reduce to
\begin{eqnarray}
  X^{\prime\prime}(x) + k^2 X(x)
  &=&
  0
  \hskip1.8cm 
  0 < x < L 
\label{eq_oneDrodAX}
\\[5pt]
  \alpha_1 X(0) + \beta_1 X^\prime(0) &=& 0
\label{eq_oneDrodBX}  
\\[-3pt]
  \alpha_2 X(L) + \beta_2 X^\prime(L) &=& 0
\nonumber
  \ ,
\end{eqnarray}
where, now, the condition $X(x)=T_0(x)$ is the obvious statement 
that $X(x)$ is simply the initial condition of the original problem. 
The general solution to (\ref{eq_oneDrodAX}) is
\begin{eqnarray}
  X_k(x)  =  A_k \cos kx + B_k \sin kx
  \ ,
  \label{Xgen}
\end{eqnarray}
and when the BC\rq{}s are applied, the modes $X_k$ will be orthogonal, 
\begin{eqnarray}
  \int_0^L dx \, X_k(x) X_{k^\prime}(x)
  =
  N_k \, \delta_{k k^\prime}
  \ .
  \label{XkXkprime}
\end{eqnarray}
Since the solutions are square integrable, and since the DE is liner and 
the BC\rq{}s are homogeneous, we have scaled $X_k$ to give an arbitrary
normalization constant $N_k$, which can be chosen for convenience. 

The general time dependent solution is a sum over all modes, 
\begin{eqnarray}
  \tilde T(x,t) = {\sum}_k  D_k \, X_k(x) \, e^{-\kappa \, k^2 t}
  \ ,
\end{eqnarray}
where we have absorbed the coefficient $U_0$ into the coefficients 
$D_k$. The $D_k$\rq{}s themselves are chosen so that the initial 
condition is satisfied,
\begin{eqnarray}
  \tilde T(x,0) &=&  {\sum}_k D_k X_k(x) =  T_0(x)
  \\[5pt]
  ~~~\Rightarrow~~~ 
    D_k &=& \frac{1}{N_k} \int_0^L \! dx\, T_0(x) \, X_k(x) 
  \label{eq_Tcoeff}
  \ .
\end{eqnarray}
For tractability, we take the IC to be linear,  as given in (\ref{IClinear}),
where $T_\smL$ is the temperature at $x=0^+$, and $T_\smR$ is the 
temperature at $x=L^-$. When $T_\smL=T_\smR$, the IC is a constant.
The linear initial condition (\ref{IClinear}) contains two temperature
parameters, $T_0(x)=T_0^\text{lin}(x; T_\smL, T_\smR)$, and therefore 
the corresponding Fourier coefficients are functions of these parameters,
\begin{eqnarray}
    D_k^\text{lin}(T_\smL, T_\smR)
    &=& 
    \frac{1}{N_k} \int_0^L \! dx\, T_0^\text{lin}(x; T_\smL, T_\smR) \, X_k(x) 
  \label{eq_TcoeffLin}
  \ .
\end{eqnarray}
When solving for the full nonhomogeneous solution (NH), rather than using
(\ref{eq_Tcoeff}) to find $D_k$, we need to choose the coefficients such that
\begin{eqnarray}
    D_k^\smNH
    &=& 
    \frac{1}{N_k} \int_0^L \! dx\, \Big[T_0(x) - \bar T(x) \Big] \,   X_k(x) 
    \\[5pt]
    &=&
    \frac{1}{N_k} \int_0^L \! dx\, \Big[T_0^\text{lin}(x; T_\smL, T_\smR) - 
    T_0^\text{lin}(x; T_1, T_2) \Big] \,   X_k(x) 
  \label{eq_DcoeffNH}
  \ ,
\end{eqnarray}
where we have written the nonhomogeneous solution $\bar T(x)$ can be 
written 
\begin{eqnarray}
  \bar T(x) 
  =
  T_0^\text{lin}(x;T_1,T_2)
  \label{barTlin}
  \ ,
\end{eqnarray}
as discussed in Section~\ref{sec:nonhomogeneous}. Therefore, the
nonhomogeneous coefficients can be expressed in terms of the homogeneous
coefficients by
\begin{eqnarray}
  D_k^\smNH(T_\smL, T_\smR, T_1, T_2)
  &=&
  D_k^\text{lin}(T_\smL - T_1, T_\smR - T_2)
  \\[5pt]
  &=&
  \frac{1}{N_k} \int_0^L \! dx\, T_0^\text{lin}(x, T_\smL - T_1, 
  T_\smR - T_2) \, X_k(x) 
  \label{eq_DcoeffNHTLTSR}
  \ .
\end{eqnarray}
We will employ this equation in the final section.

It is instructive to prove the orthogonality relation  (\ref{XkXkprime})
directly from the differential equation. To see this, multiply 
(\ref{eq_oneDrodAX}) by $X_{k^\prime}$, and then write the result 
in the two alternate forms, 
\begin{eqnarray}
  X_{k^\prime}\Big[ X_k^{\prime\prime} + k^2 X_k \Big] &=& 0
  \\[5pt]
  X_k \Big[ X_{k^\prime}^{\prime\prime} + k^{\prime\, 2} X_{k^\prime} \Big] 
  &=& 0  
 \ .
\end{eqnarray}
Upon subtracting these equations, and then integrating over space, 
we find
\begin{eqnarray}
  (k^2 - k^{\prime \, 2})  
  \int_0^L \! dx \,  X_k \, X_{k^\prime}
  &=&
  \int_0^L \! dx \, 
  \Big[ X_k X_{k^\prime}^{\prime\prime} - 
   X_{k^\prime} X_k^{\prime\prime} 
  \Big]
  \\[5pt]
  \nonumber
  &=&
  \int_0^L \! dx \, 
  \Big[ \frac{d}{dx}\,\Big(X_k X_{k^\prime}^{\prime}\Big) 
  -
  X_k^\prime  X_{k^\prime}^\prime
  - 
  \frac{d}{dx}\Big( X_{k^\prime} X_k^{\prime}  \Big)
  +
  X_{k^\prime}^\prime X_k^{\prime}
  \Big]
  \\[5pt]
  &=&
  \int_0^L \! dx \, 
  \frac{d}{dx}\,\Big(X_k X_{k^\prime}^{\prime}
  - 
   X_{k^\prime} X_k^{\prime}  \Big)
   \\[5pt]
   &=&
  \Big(X_k X_{k^\prime}^{\prime}
   - 
   X_{k^\prime} X_k^{\prime}  \Big)
   \Big\vert_0^L
   =
   0
   \ ,
\end{eqnarray}
where each contribution from $x=0$ and $x=L$ vanishes separately
because of their respective boundary conditions. We therefore arrive at
\begin{eqnarray}
  (k^2 - k^{\prime \, 2})  
  \int_0^L \! dx \,  X_k \, X_{k^\prime} = 0
  \ .
  \label{eq_kminuskXX}
\end{eqnarray}
Provided $k \ne k^\prime$, we can divide (\ref{eq_kminuskXX}) by 
$k^2 - k^{\prime \, 2}$ to obtain
\begin{eqnarray}
  \int_0^L \! dx \,  X_k(x) \, X_{k^\prime}(x)
   &=&
   0
   ~~~{\rm when}~ k \ne k^\prime
  \ .
\end{eqnarray}
However, when $k=k^\prime$, (\ref{eq_kminuskXX}) gives no constraint 
on the corresponding normalization integral; however, since the BC\rq{}s 
are homogeneous, we are free to normalize $X_k$ over $[0,L]$ such that 
$\int dx \, X_k^2 = N_k$,  for any convenient choice of $N_k$.

\subsection{Special Cases of the Homogeneous Problem}

We now find the homogeneous solutions for four special boundary
conditions, BC1--BC4.

\subsubsection{BC1}

\begin{figure}[t!]
\includegraphics[scale=0.45]{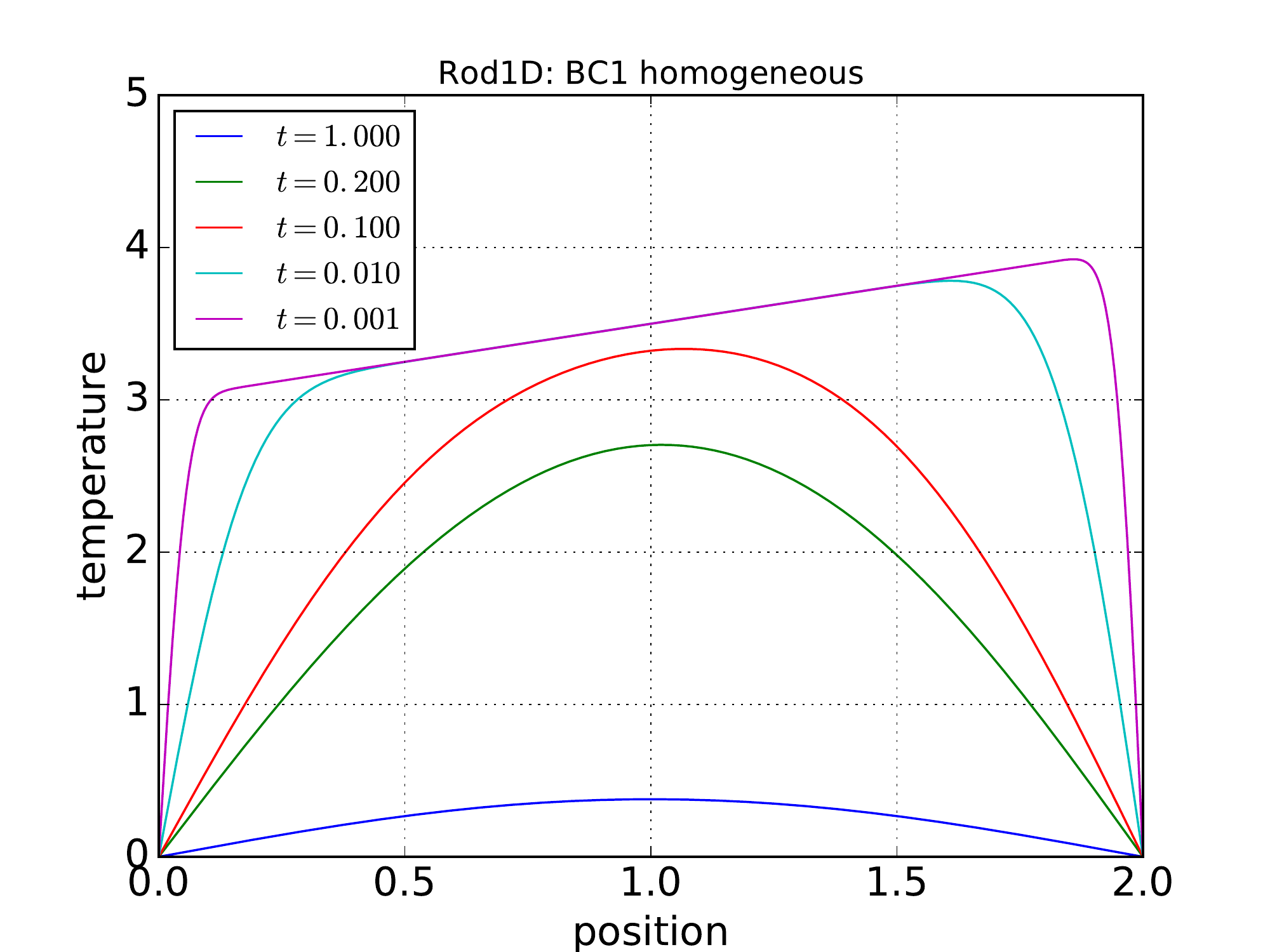} 
\caption{\footnoteskip  
This is the same as Fig.~\ref{fig_planar_sandwich_homo_ep},
the homogeneous planar sandwich, except we use the base 
class Rod1D(alpha1=1, beta1=0, alpha2=1, beta2=0, TL=3, TR=4).
}
\label{fig_rod1D_BC1}
\end{figure}

The first case holds the temperature fixed to zero at both ends 
of the rod,
\begin{eqnarray}
  \tilde T(0,t) &=&0
  \label{BC1A}
  \\
  \tilde T(L,t) &=& 0
  \label{eq_BC1B}
  \ . 
\end{eqnarray}
The general solution $X_k(x)  \!=\!   A_k \cos kx + B_k \sin kx$ 
reduces to $X_k(x) = B_k\sin k x$ under (\ref{BC1A}), while (\ref{eq_BC1B}) 
restricts the wave numbers
to satisfy $\sin k L = 0$, {\em i.e.} $k=k_n =n \pi/L$ for $n = 1, 2, 3, \cdots$.
Note that $n=0$ does not contribute, since this gives the trivial vanishing
solution. It is convenient to express the modes by $X_n(x)=\sin k_n x$,
separating the coefficient $B_n = B_{k_n}$ from the mode $X_n$ itself. The
homogeneous solution then takes the form
\begin{eqnarray}
 \tilde T(x,t)
 &=&
 \sum_{n=1}^\infty 
 B_n \, X_n(x)\, e^{-\kappa \, k_n^2 t}
 \\[5pt]
  X_n(x) &=& \sin k_n x 
  \\[5pt]
  k_n &=& 
  \frac{n \pi}{L}
    \hskip1.0cm    n = 1, 2, 3, \cdots  
  \ .
\end{eqnarray}
The tilde over the temperature is meant to explicitly remind us that this 
is the general {\em homogeneous} solution. The orthogonality condition 
on the modes $X_n$ can be checked by a simple integration,
\begin{eqnarray}
 \int_0^L dx \, X_n(x) X_m(x) &=& \frac{L}{2}\, \delta_{nm}
 \ .
\end{eqnarray}
For an initial condition $\tilde T(x,0) =T_0(x)$, we can calculate 
the corresponding coefficients in the Fourier sum,
\begin{eqnarray}
  B_n 
  = 
  \frac{2}{L}
  \int_0^L dx\, T_0(x) \sin k_n x 
  \ .
\end{eqnarray}
For the linear initial condition (\ref{IClinear}), a simple calculation gives
\begin{eqnarray}
  B_n 
  &=& 
  2 T_\smL \, \frac{1 - (-1)^n}{n\pi}
  +
  2(T_\smL - T_\smR) \,\frac{(-1)^n}{n \pi}
  \label{eq_Bnfirstline}
  \\[8pt]
  &=&
  \frac{2 T_\smL - 2 T_\smR (-1)^n}{n\pi}
  \label{eq_Bnsecdondline}
  \ .
\end{eqnarray}
The first two terms in line (\ref{eq_Bnfirstline}) are the constant and 
linear contributions of $T_0(x)$, respectively, and a typical solution 
is illustrated in Fig.~\ref{fig_rod1D_BC1}. 
The ExactPack object used to create Fig.~\ref{fig_rod1D_BC1} is
the class \verb+Rod1D+, which takes the following boundary and
initial condition arguments

\vskip0.2cm
\noindent
\verb+Rod1D(alpha1=1, beta1=0, alpha2=1, beta2=0, TL=3, TR=4)+ \ .
\vskip0.2cm

\noindent
This Figure is identical to Fig.~\ref{fig_planar_sandwich_homo_ep}, and 
is meant to illustrate the parent class Rod1D from which PlanarSandwich 
inherits.

\subsubsection{BC2}

The second special boundary condition that we consider sets the heat 
flux at both ends of the rod to zero,
\begin{eqnarray}
  \partial_x \tilde T(0,t) &=& 0 
  \label{eq_BC2A}
  \\
  \partial_x \tilde T(L,t) &=& 0 
  \label{eq_BC2B}
  \ .
\end{eqnarray}
This is the hot planar sandwich of the introduction. The general 
solution $X_k(x)  \!=\!   A_k \cos kx + B_k \sin kx$ reduces to 
$X_k(x) = A_k \cos k x$ under (\ref{eq_BC2A}) , while (\ref{eq_BC2B}) 
restricts the wave numbers to 
$k \sin k L = 0$,  so that $k=k_n=n \pi/L$ for $n=0,1,2 \cdots$. In 
this case, the $n=0$ mode is permitted (and essential). As before we 
separate the Fourier coefficients $A_n = A_{k_n}$ from the mode 
functions themselves, $X_n=X_{k_n}$, and we write
\begin{eqnarray}
  \tilde T(x,t)  &=& 
  \frac{A_0}{2} + \sum_{n=1}^\infty A_n \, X_n(x)\,
  e^{-\kappa\, k_n^2 t}
  \label{A0half}
   \\[5pt]
  X_n(x) &=& \cos k_n x   
  \\[5pt]
    k_n &=& \frac{n \pi}{L}
  \hskip1.0cm
   n = 0, 1, 2, \cdots
  \ .
\end{eqnarray}
A conventional factor of $1/2$ has been used in the $n=0$ term because 
of the  difference in normalization between $n=0$ and $n \ne 0$,
\begin{eqnarray}
  \int_0^L dx \, X_0^2(x) &=& L
  \\[5pt]
  \int_0^L dx \, X_n^2(x) &=& \frac{L}{2} ~~~~ n \ne 0
  \ ,
\end{eqnarray}
since $X_0(x)=1$ and $X_n = \cos k_n x$. Given the initial condition 
$\tilde T(x,0) = T_0(x)$, the Fourier modes become
\begin{eqnarray}
  A_n = \frac{2}{L}\, \int_0^L dx \, T_0(x) \cos k_n x
  \label{AnBCTwo}
  \ .
\end{eqnarray}
This holds for all values of $n$, including $n=0$, because we have
inserted the factor of 1/2 in the $A_0$-term of (\ref{A0half}).
For simplicity, we will take the linear initial condition (\ref{IClinear})
for $T_0(x)$, in which case,  (\ref{AnBCTwo}) gives the coefficients
\begin{eqnarray}
  \frac{A_0}{2}
  &=& 
  \frac{1}{2}\Big(T_\smL + T_\smR\Big)
  \label{eq_Azeroovertwo}
  \\[5pt]
  A_n 
  &=& 
  2\, \Big(T_\smL - T_\smR \Big) \, \frac{1 - (-1)^n}{n^2 \pi^2}
  \label{eq_Annotzero}
  \ .
\end{eqnarray}
For pedagogical purposes, let us be pedantic and work through 
the algebra for the $A_n$ coefficients, doing the $n=0$ case first:
\begin{eqnarray}
  \frac{A_0}{2}
  &=& 
  \frac{1}{L} \int_0^L \, T_0(x)
  =
  \frac{1}{L} \int_0^L \, \left[
    T_\smL + \frac{T_\smR - T_\smL}{L}\, x
    \right]
    \\[5pt]
    &=&
    T_\sm L + \left[\frac{T_\smR - T_\smL}{2}
    \right]
    =
    \frac{1}{2}\left[ T_\smR + T_\smL \right]
    \ .
\end{eqnarray}
\begin{figure}[t!]
\includegraphics[scale=0.45]{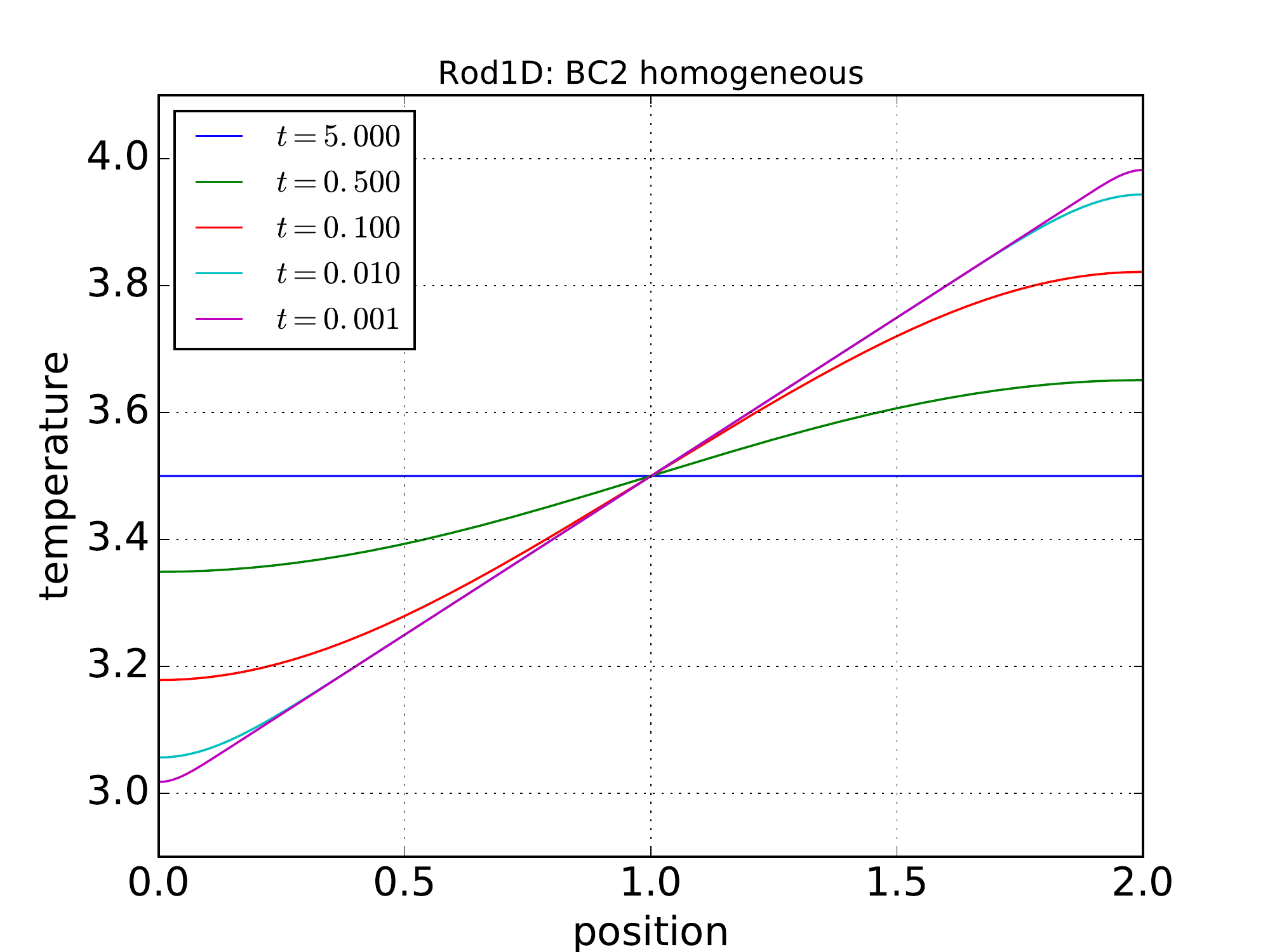} 
\caption{\footnoteskip  
  BC2 with $\kappa=1$, $L=2$, $T_\smL=3$, $T_\smR=4$.
  Rod1D(alpha1=0, beta1=1, alpha2=0, beta2=1, TL=3, TR=4).
}
\label{fig_rod1D_BC2}
\end{figure}
Next, taking $n \ne 0$, we find:
\begin{eqnarray}
  A_n
  &=& 
  \frac{2}{L} \int_0^L dx\, T_0(x) \cos k_n x
  \\[5pt]
  &=& 
    \frac{2}{L} \int_0^L dx\, \left[
    T_\smL + \frac{T_\smR - T_\smL}{L}\, x
    \right] \cos k_n x
    \\[5pt]
      &=& 
    T_\smL  \, \frac{2}{L} \int_0^L dx\, \cos k_n x
    + 
    \Big( T_\smR - T_\smL \Big) \frac{2}{L^2} \int_0^L dx \, x\, \cos k_n x
    \ .
\end{eqnarray}
The first term integrates to zero since
\begin{eqnarray}
  \frac{2}{L} \int_0^L dx \,\cos k_n x
  &=&
  \frac{2}{L} \, \sin k_n x \Big\vert_{x=0}^{x=L}
  =
  0 
  \ ,
\end{eqnarray}
and the second term gives
\begin{eqnarray}
  \frac{2}{L^2} \int_0^L dx \, x\, \cos k_n x
  &=&
  \frac{2}{L^2} \left[ \frac{\cos k_n x}{k_n^2} + \frac{x \sin k_n x}{k_n}
  \right]_{x=0}^{x=L}
  \\[5pt]
  &=&
  \frac{2}{L^2} \, \frac{L^2}{n^2 \pi^2}\,\Big[\cos k_n L - 1 \Big]
  =
  2 \, \frac{(-1)^n - 1}{n^2 \pi^2}
  \ ,
\end{eqnarray}
which leads to (\ref{eq_Annotzero}).

\subsubsection{BC3}

The next specialized boundary condition is
\begin{eqnarray}
  \tilde T(0,t) &=&0
  \label{eq_BC3A}
  \\
  \partial_x \tilde T(L,t) &=& 0
  \label{eq_BC3B}
  \ .
\end{eqnarray}
The general solution $X_k(x)  \!=\!   A_k \cos kx + B_k \sin kx$ 
under (\ref{eq_BC3A}) reduces to
$X_k(x) = B_k \sin k x$, while (\ref{eq_BC3B}) restricts the wave 
numbers to $k \cos k L = 0$, so that $k=k_n=(2 n + 1) \pi/2L$ for 
$n=0,1,2 \cdots$.  The general homogeneous solution is therefore
\begin{eqnarray}
  \tilde T(x,t)  &=& \sum_{n=0}^\infty 
  B_n \, X_n(x) \, e^{-\kappa \, k_n^2 t}
  \\[5pt]
  X_n(x) &=& \sin k_n x  
  \\[5pt]
    k_n &=& 
  \frac{(2 n + 1) \pi}{2 L}
  \hskip1.0cm
   n = 0, 1, 2, \cdots
  \ .
\end{eqnarray}
The initial condition $\tilde T(x,0) = T_0(x)$ gives the Fourier modes
\begin{eqnarray}
  B_n = \frac{2}{L}\, \int_0^L dx \, T_0(x) \sin k_n x
  \ ,
  \label{eq_Bnbcthree}  
\end{eqnarray}
and, as before, upon taking the linear function (\ref{IClinear}),  we find
\begin{eqnarray}
  B_n 
  &=&  
  \frac{4 T_\smL}{(2 n + 1) \pi}
  +
  4\big(T_\smR - T_\smL\big)\left[
  \frac{1}{(2n+1) \pi} 
  -
  \frac{2}{(2n+1)^2 \pi^2} 
  \right]
  \\[5pt]
    &=&  
  \frac{4 T_\smR}{(2 n + 1) \pi}
  -
  \frac{8\big(T_\smR - T_\smL\big) }{(2n+1)^2 \pi^2} 
  \ .
\end{eqnarray}
Before plotting this example, let us examine the next boundary condition.

\subsubsection{BC4}

\begin{figure}[t!]
\includegraphics[scale=0.45]{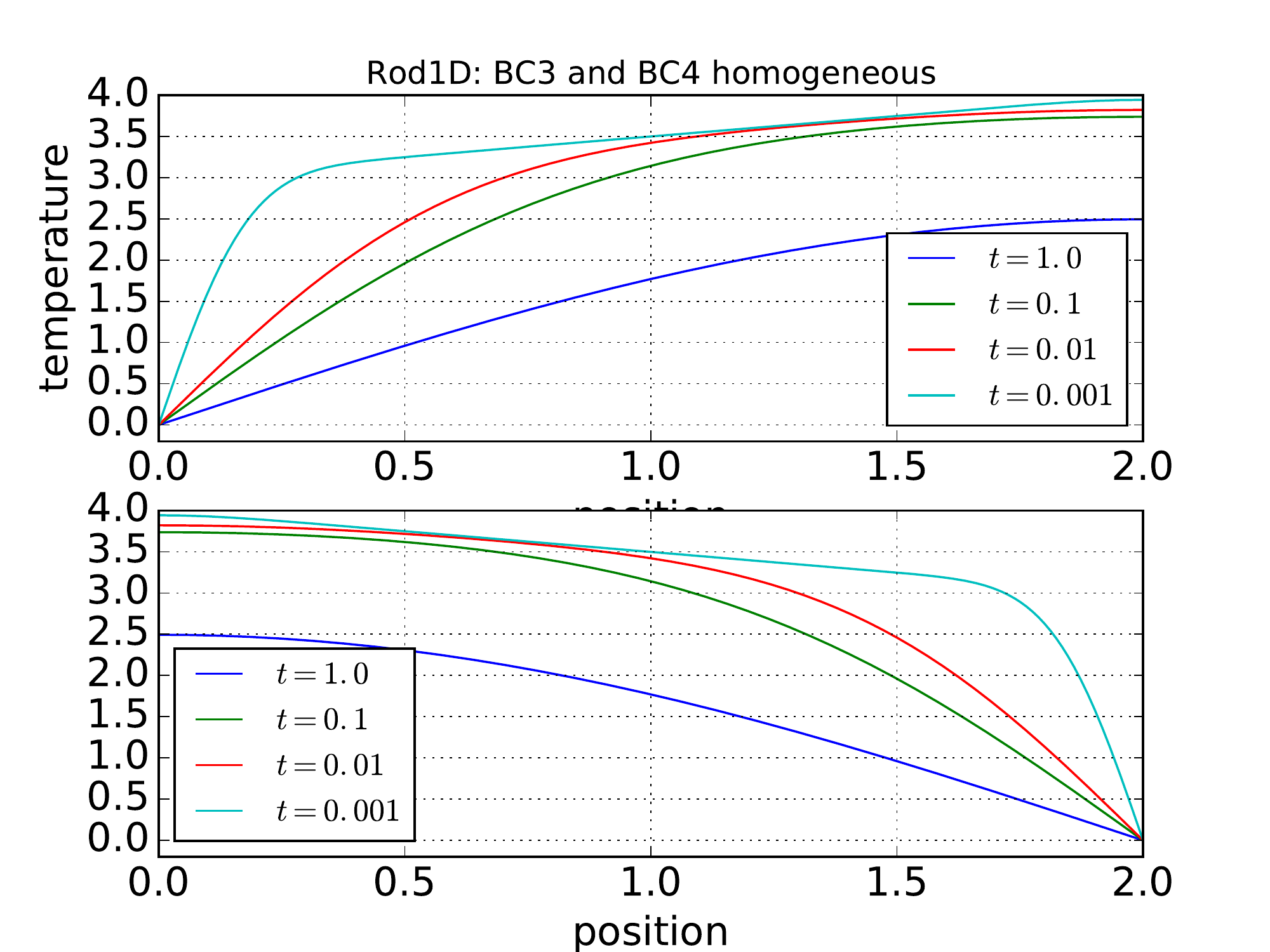} 
\caption{\footnoteskip  
BC3  and BC4 for $\kappa=1$, $L=2$, $T_\smL=T_\smR=3$.  By 
symmetry principles, the two profiles are mirror images of one another. 
BC3 is instantiated by Rod1D(alpha1=1, beta1=0, alpha2=0, beta2=1, 
TL=3, TR=4), and BC4 by Rod1D(alpha1=0, beta1=1, alpha2=1, beta2=0, 
TL=4, TR=3). Note that $T_\smL$ and $T_\smR$ are interchanged between
BC3 and BC4.
}
\label{fig_rod1D_BC3_BC4}
\end{figure}

The last special case is the boundary condition
\begin{eqnarray}
  \partial_x \tilde T(0,t) &=& 0 
  \label{eq_BC4A}
  \\
  \tilde T(L,t) &=& 0
  \ .
 \label{eq_BC4B}
\end{eqnarray}
The general solution $X_k(x) \!=\!  A_k \cos kx + B_k \sin kx$ 
reduces to $X_k(x) = A_k \cos k x $ under (\ref{eq_BC3A}), 
while (\ref{eq_BC4B}) restricts the wave 
numbers to $\cos k L = 0$, {\em i.e.} $k=k_n=(2 n + 1) \pi/2L$ for 
$n=0,1,2 \cdots$, which gives rise to the homogeneous solution
\begin{eqnarray}
  \tilde T(x,t)  &=& \sum_{n=0}^\infty 
  A_n \, X_n(x)\, e^{-\kappa \, k_n^2 t}
  \\[5pt]
  X_n(x) &=& \cos k_n x  
  \\[5pt]
    k_n &=& 
  \frac{(2 n + 1) \pi}{2 L}
  \hskip1.0cm 
  n = 0, 1, 2, \cdots
  \ .
\end{eqnarray}
Similar to (\ref{eq_Bnbcthree}), the mode coefficient is
\begin{eqnarray}
  A_n = \frac{2}{L}\, \int_0^L dx \, T_0(x) \cos k_n x
  \ ,
\end{eqnarray}
and, upon taking the linear initial condition (\ref{IClinear}), we find
\begin{eqnarray}
  A_n 
  &=&
  4 T_\smL \, \frac{(-1)^n }{(2 n + 1)\pi }
  -
  8\Big(T_\smR - T_\smL \Big) \,\frac{1 - (-1)^n}{(2n + 1)^2\, \pi^2}
  \ .
\end{eqnarray}
The cases BC3 and BC4 are plotted in Fig.~\ref{fig_rod1D_BC3_BC4}.

\subsection{General Boundary Conditions}

\begin{figure}[h!]
\includegraphics[scale=0.45]{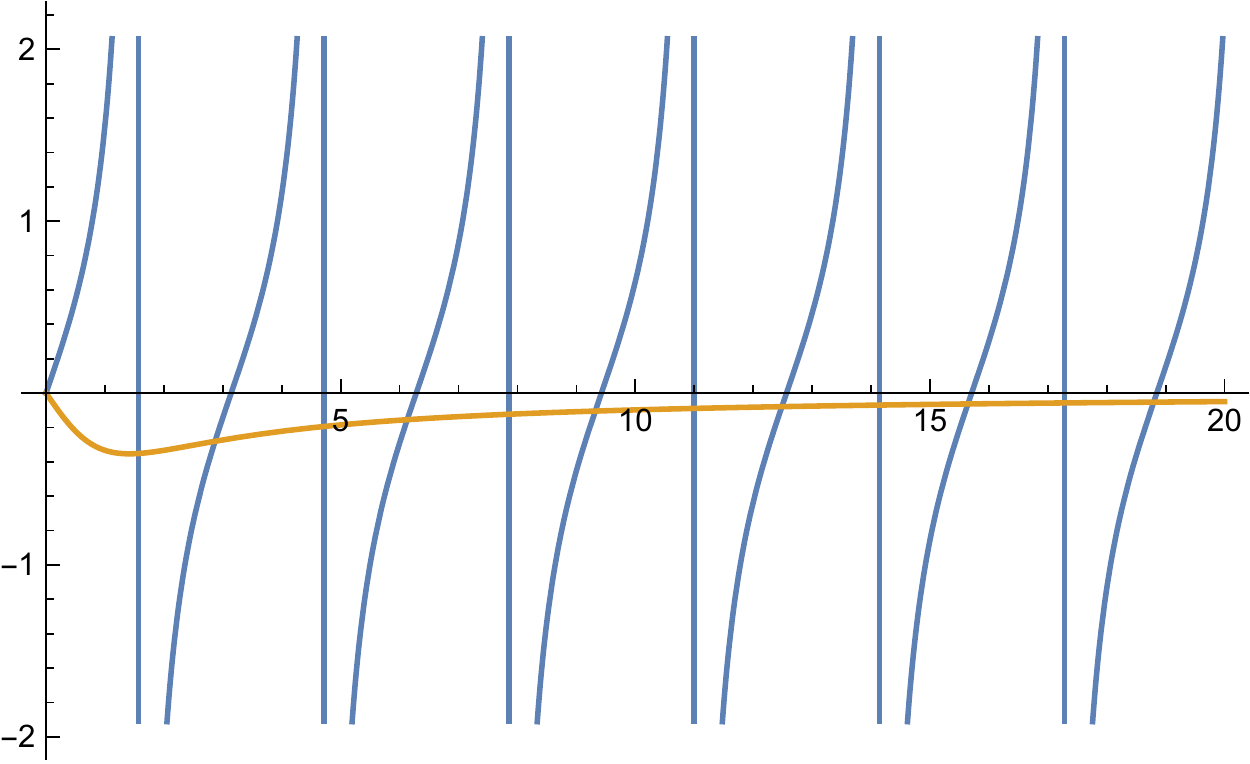} 
\caption{\footnoteskip  
  The roots $\mu_n$ for $\alpha_1=1$, $\bar\beta=1/2$, $\alpha_2=1$, and
  $\bar\beta_2=1$. For $L=2$ this gives $\beta_1=1$ and $\beta_2=2$.
}
\label{fig_tankL}
\end{figure}

We now turn to the general form of the boundary conditions,
which, expressed in terms of $X$, take the form
\begin{eqnarray}
  \alpha_1 X_k(0) + \beta_1 X_k^\prime(0) &=& 0
\label{eq_XBCzero}  
\\
  \alpha_2 X_k(L) + \beta_2 X_k^\prime(L)&=& 0
  \ .
\label{eq_XBCL}  
\end{eqnarray}
The solution and its derivative are
\begin{eqnarray}
  X_k(x) &=& A \cos k x + B \sin k x
  \\
  X_k^\prime(x) &=& - A k \sin k x  + B k \cos k x
  \ .
\end{eqnarray}
Substituting this into (\ref{eq_XBCzero}) and (\ref{eq_XBCL}) gives
\begin{eqnarray}
  \alpha_1 A + \beta_1 B k &=& 0
  \label{BCa}
  \\
  \alpha_2 \Big[A \cos k L + B \sin k L \Big] 
  + 
  \beta_2\Big[ -A k \sin k L + B k \cos k L \Big] &=& 0
  \label{BCb}
  \ .
\end{eqnarray}
Upon diving by $\cos k L \ne 0$, can write (\ref{BCb}) as
\begin{eqnarray}
  (\alpha_2 \, B    -\beta_2 \, A k) \tan k L 
  + \alpha_2 A + \beta_2 \, B k  &=& 0
  \ ,
\end{eqnarray}
or
\begin{eqnarray}
   \tan k L 
   &=&
   \frac{\beta_2 \, B k + \alpha_2 A }{ \beta_2 \, A k - \alpha_2 \, B }
   \ .
   \label{BCtankL}
\end{eqnarray}
From (\ref{BCa}) we have $B k = - \alpha_1 A / \beta_1$ (if $\beta_1 \ne 0)$, 
and substituting into (\ref{BCtankL}) gives
\begin{eqnarray}
   \tan k L 
   &=&
   \frac{-(\alpha_1 \beta_2 /\beta_1) + \alpha_2 }{ \beta_2 \, k +  \alpha_2 \,
    (\alpha_1 / \beta_1 k) }   
   \cdot 
   \frac{\beta_1 k}{\beta_1 k}
   \\[5pt]
   &=&
   \frac{-\alpha_1 \beta_2 \, k + \alpha_2 \beta_1 k }{ \beta_1 \beta_2 \, k^2 
   +  \alpha_2 \, \alpha_1  }      
  \ .
  \label{BCtankLone}
\end{eqnarray}
Setting $\mu \equiv k L$ and $\bar\beta_i \equiv \beta_i/L$, we can write (\ref{BCtankLone}) in the form
\begin{eqnarray}
   \tan \mu
   &=&
   \frac{ (\alpha_2 \bar\beta_1 - \alpha_1 \bar\beta_2) \, \mu }{\alpha_1 \alpha_2  +  \bar\beta_1 \bar\beta_2\, \mu^2  }      
   \ .
  \label{BCtanmu}
\end{eqnarray}
The solution is illustrated in Fig.~\ref{fig_tankL}.
Equation (\ref{BCtanmu}) will give solutions $\mu_n$ for $n=0,1,2,\cdots$ 
and with wave numbers
\begin{eqnarray}
  k_n = \frac{\mu_n}{L}
  \ .
  \label{knDef}
\end{eqnarray}
Note that $\mu_0=0$, and therefore $k_0=0$. The solution now takes the form
\begin{eqnarray}
  X_n(x) &=& A_n \cos k_n x + B_n \sin k_n
  \label{SolAB}
  \\[5pt]
  A_n &=& - \frac{\beta_1 k_n}{\alpha_1}\, B_n
  \ ,
\end{eqnarray}
where $\alpha_1 \ne 0$. The case of $\alpha_1=0$ will be handled 
separately. Setting $B_n=1$ for convenient, the solution (\ref{SolAB}) 
can be expressed as
\begin{eqnarray}
  X_n(x) &=& \sin k_n x - \frac{\beta_1 k_n}{\alpha_1} \cos k_n x  
  \ .
  \label{SolABtwo}
\end{eqnarray}
And the general solution is 
\begin{eqnarray}
  X(x) &=& \sum_{n=1}^\infty B_n X_n(x)
  \ ,
\end{eqnarray}
as the $n=0$ term does not contribute. Note that 
\begin{eqnarray}
  \int_0^L dx \, X_n(x) X_m(x) = 0 ~~~{\rm for}~ n \ne m 
  \label{SolABtwo}
\end{eqnarray}
and
\begin{eqnarray}
  \int_0^L dx \, X_n^2(x)  
  &=&
  \frac{1}{4 k_n \alpha_1^2}\Bigg[
  -2 \alpha_1 \beta_1 k_n  + 2 (\beta_1^2 k_n^2 + \alpha_1^2 ) k_nL +
  \\
  \nonumber && \hskip1.5cm
  2\alpha_1 \beta_1 k_n \cos 2 k_n L + 
  (\beta_1^2 k_n^2 - \alpha_1^2 ) \sin 2 k_n L
  \Bigg]
  \ . 
  \label{SolABthree}
\end{eqnarray}
In summary,
\begin{eqnarray}
 &&
  \int_0^L dx \, X_n(x) X_m(x) 
  =
  N_n \, \delta_{nm} \ ,
  \\
  &&N_n
  =
  \frac{1}{4 k_n \alpha_1^2}\Big[
  -2 \alpha_1 \beta_1 k_n  + 2 (\beta_1^2 k_n^2 + \alpha_1^2 ) k_nL +
  2\alpha_1 \beta_1 k_n \cos 2 k_n L + 
  (\beta_1^2 k_n^2 - \alpha_1^2 ) \sin 2 k_n L
  \Big] \ .
  \nonumber
  \\
  \label{SolABfour}
\end{eqnarray}
Since $k_0=0$, we have $X_0(x)=0$, so we are free to restrict $n=1,2,3,\cdots$,
and the general solution is
\begin{eqnarray}
  X(x) 
  &=& \sum_{n=1}^\infty D_n\, X_n(x)
  \ .
\end{eqnarray}
Since $X(x)=T_0(x)$, we find
\begin{eqnarray}
  D_n = \frac{1}{N_n}\int_0^L dx\, T_0(x) X_n(x)
  \ .
\end{eqnarray}
It is convenient for numerical work to express this in
terms of $A_n$ and $B_n$ coefficients:
\begin{eqnarray}
  X(x) 
  &=&
  \sum_{n=1}^\infty D_n\, \Big[ 
  - \frac{\beta_1 k_n}{\alpha_1}\,\cos k_n x + \sin k_n x 
  \Big]
  \\[5pt]
  &=&
  \sum_{n=1}^\infty\Big[ 
  A_n \cos k_n x + B_n \sin k_n x 
  \Big]
  {\rm ~~~with}
  \\[5pt]
  \nonumber
  A_n &=& - \frac{\beta_1 k_n}{\alpha_1}\, D_n
  \\[5pt]
  \nonumber
  B_n &=& D_n
  \ .
\end{eqnarray}
The temperature $\tilde T(x,t)$ is therefore,
\begin{eqnarray}
  \tilde T(x,t) 
  &=&
  \sum_{n=1}^\infty\Big[ 
  A_n \cos k_n x + B_n \sin k_n x 
  \Big] \, e^{-\kappa \, k_n^2 t}
  \\[5pt]
    B_n &=& \frac{1}{N_n}\int_0^L dx\, T_0(x) X_n(x)
  \\[5pt]
  A_n &=& - \frac{\beta_1 k_n}{\alpha_1}\, B_n
  \ .
\end{eqnarray}
For $T_0^a(x)=T_1$ we have
\begin{eqnarray}
  B_n^a &=& \frac{T_1}{N_n} \, \left[
  \frac{1 - \cos k_n L}{k_n}  - \frac{\beta_1 \sin k_n L}{\alpha_1}\right]
  \ .
\end{eqnarray}
For $T_0^b(x) = (T_2 - T_1) \, x/L$ we have
\begin{eqnarray}
  B_n^b &=& \frac{T_2 - T_1}{N_n\, L}\,\frac{1}{\alpha_1 k_n^2} \, \Big[
  \beta_1 k_n  - (\alpha_1 k_n L + \beta_1 k_n) \cos k_n L + (\alpha_1 - 
  \beta_1 k_n^2 L) \sin k_n L 
  \Big]
  \ ,
\end{eqnarray}
with $B_n = B_n^a + B_n^b$.

\begin{figure}[h!]
\includegraphics[scale=0.45]{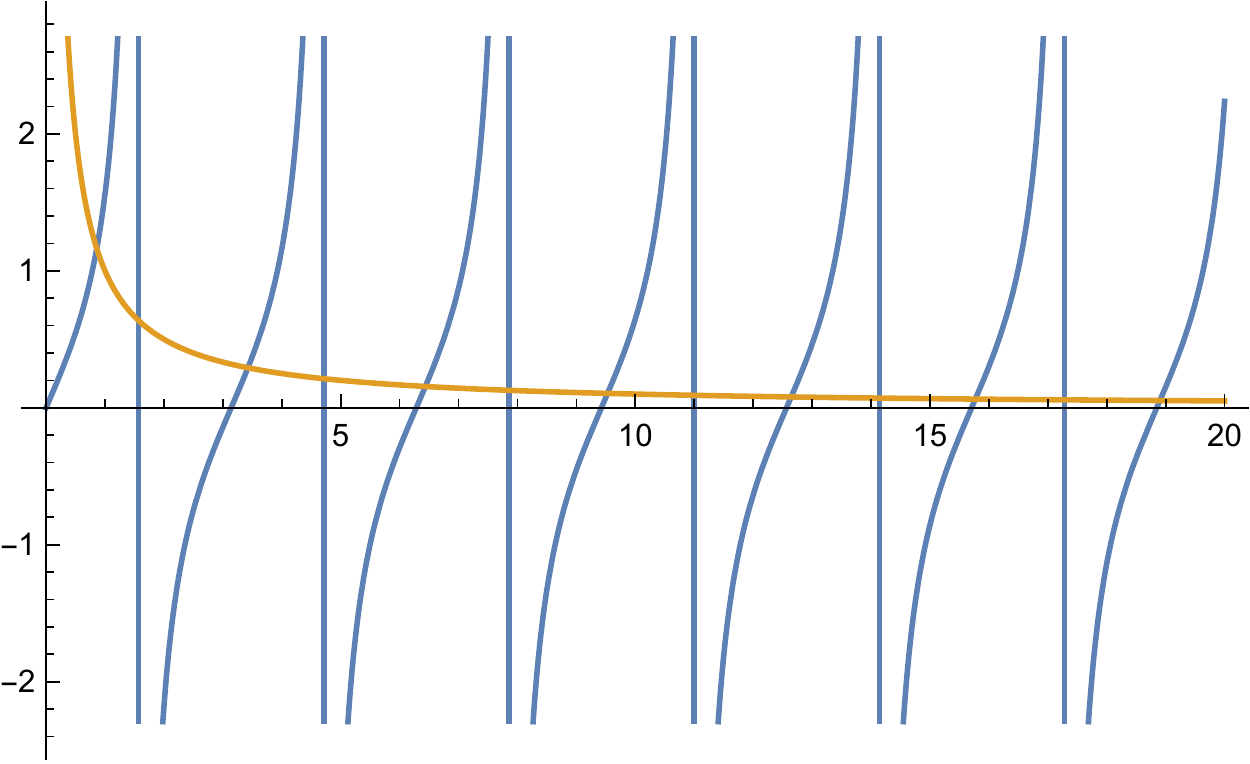} 
\caption{\footnoteskip  
  The roots $\mu_n$ for $\alpha_1=0$, $\alpha_2=1$, and
  $\bar\beta_2=1$. For $L=2$ we have $\beta_2=2$.
}
\label{fig_tankLzero}
\end{figure}
Let us now consider the case of $\alpha_1=0$, so that (\ref{BCtanmu}) becomes
\begin{eqnarray}
   \tan \mu
   &=&
   \frac{a}{\mu}
   ~~~~{\rm with} ~~    a = \alpha_2 / \bar\beta_2
   \ .
  \label{BCtanmuzero}
\end{eqnarray}
We can find an approximate solution for large values of $\mu$: since
the RHS is very small for $\mu \gg 1$, we must solve $\tan \mu = 0$,
and therefore $\mu_n^{(0)} = n \pi$. The exact solution can be expressed
as $\mu_n = n \pi + h$, where $h$ is small and unknown. Then ${\rm LHS} =
\tan(n\pi + h) = \tan(h) = h + {\cal O}(h^2)$. Similarly, ${\rm RHS}=
a/(n\pi + h) = (a/n\pi)\big(1 + h /n\pi \big)^{-1} =  (a/n\pi)\big(1 - h /n\pi
\big) + {\cal O}([h/n]^2) = a/n\pi - a h + {\cal O}([h/n]^2)$, thus
\begin{eqnarray}
  h = \frac{a}{n\pi} - a h 
  ~~~\Rightarrow~~~
  h = \frac{a}{1 + a}\,\frac{1}{n\pi}
  \ ,
\end{eqnarray}
and the first order solution becomes
\begin{eqnarray}
  \mu_n^{(1)}
  =
  n \pi + \frac{a}{1 + a}\,\frac{1}{n\pi} + {\cal O}(1/n^2)
  \ .
\end{eqnarray}
This can be used as an initial guess when using an iteration
method to find the $\mu_n$. The solution is
\begin{eqnarray}
  T(x,t) &=& \sum_{n=1}^\infty A_n X_n(x)\,e^{-\kappa \, k_n^2 t}
  \\
  X_n(x) &=& \cos k_n L
  \\
  \int_0^L dx \, X_n(x) X_m(x) &=& N_n \, \delta_{nm}
 \\
  N_n &=& \frac{1}{4 k_n}\,\Big[2 k_n L + \sin 2 k_n L \Big]
  \ ,
\end{eqnarray}
and
\begin{eqnarray}
  A_n 
  &=&
  \frac{1}{N_n}\int_0^L dx\, T(x,0) X_n(x)
  \\[5pt]
  &=&
  \frac{T_1}{k_n} \,\sin k_n L + \frac{T_2 - T_1}{k_n^2 L}\,
  \Big[ -1 + \cos k_n L + k_n L \sin k_n L \Big]
  \ .
\end{eqnarray}
\pagebreak
\section{The Full Nonhomogeneous Problem}

Suppose now that $\tilde T(x,t)$ is a general solution to the homogeneous 
problem as described in the previous section. Also suppose that $\bar T(x)$ 
is a specific solution to the nonhomogeneous problem as described in the 
previous section, then
\begin{eqnarray}
  T(x,t) 
  &=& 
  \tilde T(x,t) + \bar T(x)
  \label{eq_Tgen}
\end{eqnarray}  
is the solution to the nonhomogeneous problem 
(\ref{eq_oneDrodAnh})--(\ref{eq_oneDrodCnh}). The general
homogeneous solution, and the specific nonhomogeneous
solution take the form
\begin{eqnarray}  
  \tilde T(x,t) 
  &=&
  \sum_n D_n \, X_n(x)\, e^{-\kappa \, k_n^2 t}
  \\
  \bar T(x)
  &=&
  T_0^\text{lin}(x; T_1, T_2)
  =
  T_1 + \frac{T_2 - T_1}{L}\, x
  \ ,
  \label{eq_tildeT}
\end{eqnarray}    
where the coefficients are chosen to satisfy the initial
condition, 
\begin{eqnarray}    
  D_n
  &=& 
  \int_0^L \Big[T_0(x) -   \bar T(x) \Big] X_n(x)
  \ ,
\end{eqnarray}
with $\bar T(x)$ given by (\ref{eq_tildeT}), and $T_0(x)$ given by
\begin{eqnarray}
  T_0(x) 
  &=&
  T_0^\text{lin}(x; T_\smL, T_\smR)
  =
  T_\smL + \frac{T_\smR - T_\smL}{L}\, x
  \ .
\end{eqnarray}
Since $T_0(x)$ and $\bar T(x)$ are of the same functional
form, we can write
\begin{eqnarray}    
  T_0(x) - \bar T(x)
  &\equiv&
  T_0^\text{lin}(x; T_a, T_b) 
  =
  T_a + \frac{T_b - T_a}{L}\, x
  \\
  T_a &=& T_\smL - T_1
  \\
  T_b &=& T_\smR - T_2
  \ ,
\end{eqnarray}
where we have expressed the parametric dependence upon temperature
explicitly in $T_0^\text{lin}$.  Therefore,
\begin{eqnarray}    
  D_n
  &=&
  D_n^\text{lin}(T_\smL - T_1, T_\smR - T_2)
  \equiv
  \int_0^L T_0^\text{lin}(x; T_\smL-T_1, T_\smR-T_2) \, X_n(x)
  \ .
\end{eqnarray}
This is why the the planar sandwich and the homogeneous planar
sandwich have such similar coefficients,
\begin{eqnarray}
  B_n^\text{planar sand}
  &=&
  \phantom{-}
  D_n^\text{lin}(T_1,T_2)
  \\
  B_n^\text{hom planar sand}
  &=&
  -D_n^\text{lin}(T_\smL,T_\smR)
  \ .
\end{eqnarray}
\subsection{Special Cases of the Nonhomogeneous Problem}

We turn now to the full set of nonhomogeneous problems for the
special cases considered in the previous section.

\subsubsection{BC1}

The complete solution for the nonhomogeneous BC\rq{}s
\begin{eqnarray}
  T(0,t) &=& T_1
  \\
  T(L,t) &=& T_2
\end{eqnarray}
is
\begin{eqnarray}
  T(x,t)
  =
  T_1 + \frac{(T_2 - T_1) x}{L}
  + 
  \sum_{n=1}^\infty B_n \, \sin k_n x \, e^{-\kappa\, k_n^2 t}
  \ .
\end{eqnarray}
Recall that these BC\rq{}s corresponds to $\beta_1=\beta_2=0$ with
and $\gamma_1/\alpha_1 = T_1$ and $\gamma_2/\alpha_2=T_2$
in Eqs.~(\ref{BCTone}) and (\ref{BCTtwo} ). In terms of the BC\rq{}s, 
we can write this as
\begin{eqnarray}
  \bar T(x) = T_1 + \frac{T_2 - T_1}{L}\, x
  \ .
\end{eqnarray}
The nonhomogeneous coefficients are found by
\begin{eqnarray}
  B_n &=& \int_0^L \Big[T_0(x) - \bar T(x) \Big] \sin k_n x
  \ .
\end{eqnarray}
Since we have taken the $T_0(x)$ to be a linear equation, as is $\bar T(x)$,
we can use the previous results for a linear initial conditions by substituting 
$T_\smL \to T_a = T_\smL - T_1$ and $T_\smR \to T_b = T_\smR - T_2$ 
into (\ref{eq_Bnsecdondline}), as explained in the previous section.  In other
words,
\begin{eqnarray}
  T_0(x) - \bar T(x)
  &=&
  T_a + \frac{T_b - T_a}{L}
  \\[5pt]
  T_a &=& T_\smL - T_1
  \\
  T_b &=& T_\smR - T_2
  \ ,
\end{eqnarray}
and the coefficients of the nonhomogeneous solution become
\begin{eqnarray}
  B_n 
  &=&
  2 T_a \, \frac{1 - (-1)^n}{n\pi}
  +
  2 (T_a - T_b) \,\frac{(-1)^n}{n \pi} 
  \\[5pt]
  &=&
  \frac{2 T_a - 2 T_b (-1)^n}{n\pi}
  \ .
\end{eqnarray}
A typical example of the solution is illustrated in Fig~\ref{fig_rod1D_BC1}. 
In this
Figure, we take the initial conditions as zero temperature, with the
$x=0$ BC to be $T_1=1$, and the $x=L$ BC to be $T_2=0$, and
we see that a heat wave moves from the left end of the rod to the
right, until the the entire rod is at temperature $\bar T(x)$. This is 
just the heat conduction physics of the planar sandwich.
\begin{figure}[h!]
\includegraphics[scale=0.45]{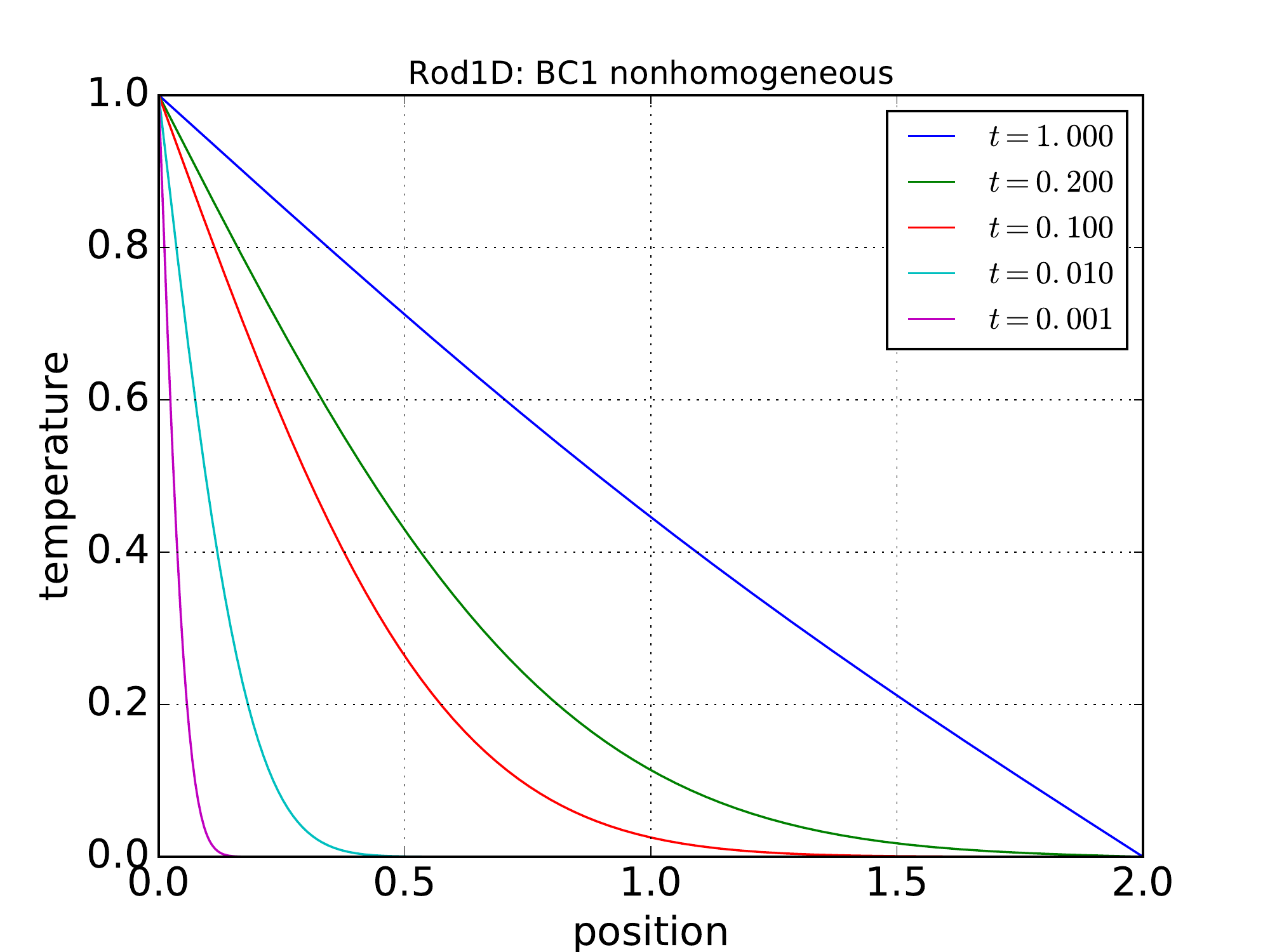} 
\caption{\footnoteskip  
  BC1 for $\kappa=1$, $L=2$,  $T_1=1$, $T_2=0$
   ($\alpha_1=1, \beta_1=0, \gamma_1=1$, and
  $\alpha_1=1, \beta_1=0, \gamma_1=0$), with
  $T_\smL=0$, $T_\smR=0$. Solver instantiation: 
  Rod1D(alpha1=1, beta1=0, alpha2=1, gamma1=1, 
  beta2=0, gamma2=0, TL=0, TR=0).
}
\label{fig_rod1D_BC1_nonhomo}
\end{figure}
\noindent
For Fig.~\ref{fig_rod1D_BC1_nonhomo}, the Class \verb+Rod1D+ takes
the boundary and initial condition arguments

\vskip0.2cm
\noindent
\verb+Rod1D(alpha1=1, beta1=0, gamma1=1, alpha2=1,beta2=0, gamma2=0, TL=0, TR=0)+.
\vskip0.2cm

\noindent
Note that $T_1=\gamma_1/\alpha_1=1$ and $T_2=\gamma_2/\alpha_2=0$.

\subsubsection{BC2}

For the boundary conditions
\begin{eqnarray}
  \partial_x T(0,t) &=& F
  \\[5pt]
  \partial_x T(L,t) &=& F
  \ ,
\end{eqnarray}
the full nonhomogeneous solution is thus
\begin{eqnarray}
  T(x,t)
  &=&
  F x + \frac{A_0}{2} + \sum_{n=1}^\infty A_n \, \cos k_n x \, 
  e^{-\kappa \, k_n^2 t}
  \ .
\end{eqnarray}
Using the initial condition $T(x,t=0)=T_0(x)$, we find
\begin{eqnarray}
  \frac{A_0}{2} + \sum_{n=1}^\infty A_n \, \cos k_n x 
  &=&
  T_0(x) - F x
  =
  T_\smL + \frac{(T_\smR - F L) - T_\smL}{L}\, x
  \ .
\end{eqnarray}
We can use the previous results (\ref{eq_Azeroovertwo}) and (\ref{eq_An})
provided we make the substitution $T_\smL \to T_a = T_\smL$ and
$T_\smR \to T_b = T_\smR - F L$,
\begin{eqnarray}
  T_a &=& T_\smL \hskip0.5cm T_b = T_\smR - F L
  \\[5pt]
  \frac{A_0}{2}
  &=& 
  \frac{1}{2}\Big(T_a + T_b\Big)
  \label{eq_Azeroovertwo}
  \\[5pt]
  A_n 
  &=& 
  2\, \Big(T_a - T_b \Big) \, \frac{1 - (-1)^n}{n^2 \pi^2}
  \label{eq_An}
  \ .
\end{eqnarray}
\begin{figure}[t!]
\includegraphics[scale=0.45]{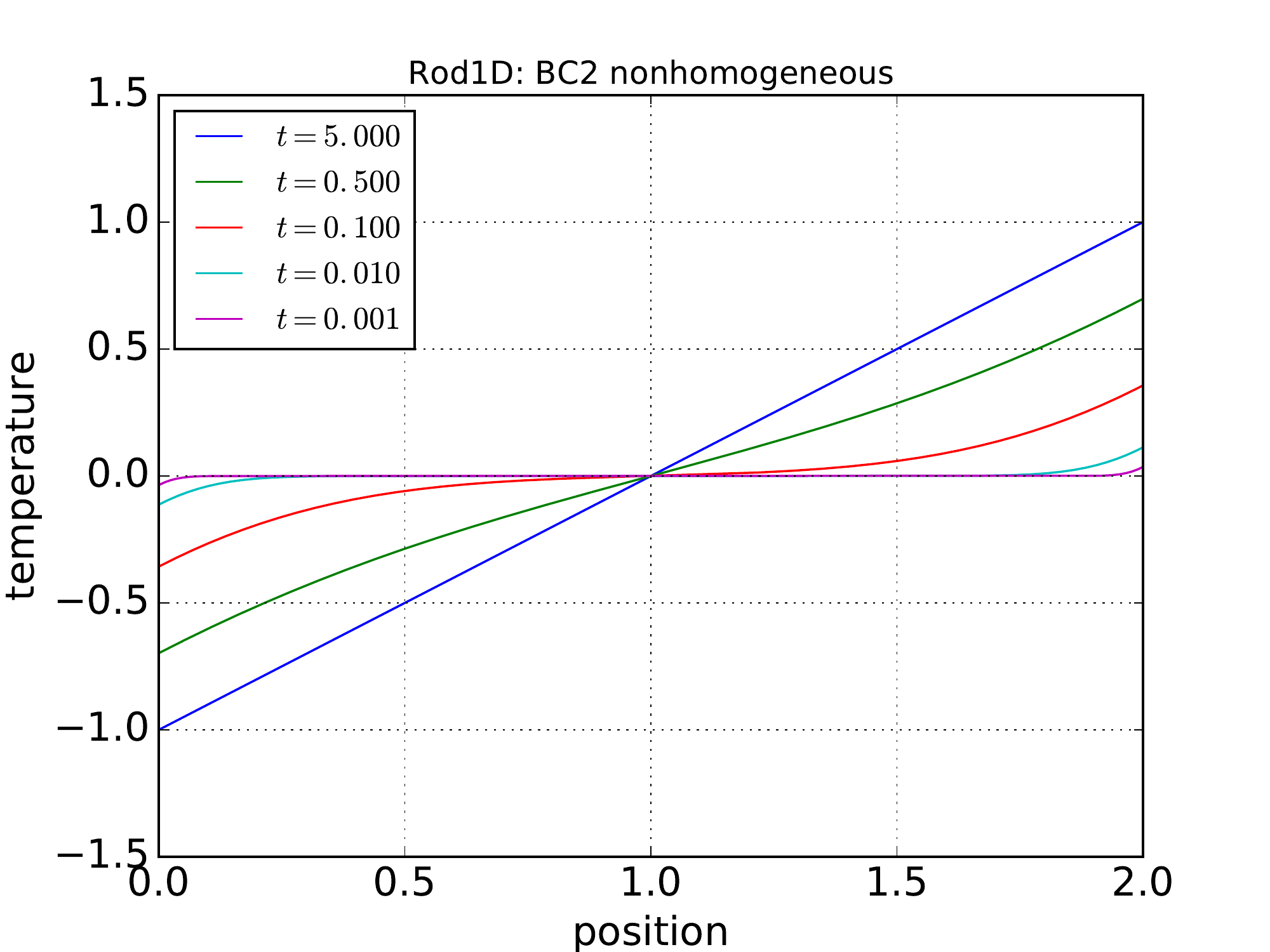} 
\caption{\footnoteskip  
  BC2 with $\kappa=1$, $L=2$, $F=1$ (with $T_\smL=0$, $T_\smR=0$).
  ExactPack instantiation: Rod1D(alpha1=0, beta1=1, gamma1=F, alpha2=0, beta2=1, gamma2=F, TL=0, TR=0).
}
\label{fig_rod1D_BC2_nonhomo}
\end{figure}
\noindent
The instantiation of \verb+Rod1D+ used for Fig.~\ref{fig_rod1D_BC2_nonhomo}
is

\vskip0.2cm
\noindent
\verb+Rod1D(alpha1=1, beta1=0, alpha2=1, gamma1=1, beta2=0, gamma2=0, TL=0, TR=0)+.
\vskip0.2cm

\noindent
Since $T_1 = \gamma_1 / \alpha_1$, and $T_2 = \gamma_2 / \alpha_2$,
we could simplify the interface to 

\vskip0.2cm
\noindent
\verb+PlanarSandwich(TL=T1, TR=T2, Nsum=1000)+.
\vskip0.2cm

\subsubsection{BC3}

For the boundary conditions
\begin{eqnarray}
  T(0,t) &=& T_1
  \\[5pt]
  \partial_x T(L,t) &=& F_2
  \ ,
\end{eqnarray}
the full nonhomogeneous solution is thus
\begin{eqnarray}
  T(x,t) &=& 
  T_1 + \frac{T_2 - T_1}{L} \, x 
  +
  \sum_{n=0}^\infty 
  B_n \, \sin k_n x \, e^{-\kappa \, k_n^2 t}
  \\[5pt]
  T_2 &=& T_1 + F_2 L
  =
  \frac{\gamma_1}{\alpha_1} + \frac{\gamma_2 L}{\beta_2}
  \\[5pt]
    k_n &=& 
  \frac{(2 n + 1) \pi}{2 L}
  \hskip1.0cm
   n = 0, 1, 2, \cdots
  \ .
\end{eqnarray}
The Fourier coefficients
\begin{eqnarray}
  B_n = \frac{2}{L}\, \int_0^L dx \, \Big[T_0(x) - \bar T(x) \Big]
  \sin k_n x
  \label{eq_Bnnonhomo}
\end{eqnarray}
take the form
\begin{eqnarray}
  B_n 
  &=&  
  \frac{4 T_a}{(2 n + 1) \pi}
  +
  4\big(T_b - T_a\big)\left[
  \frac{1}{(2n+1) \pi} 
  -
  \frac{2}{(2n+1)^2 \pi^2} 
  \right]
  \\[5pt]
    &=&  
  \frac{4 T_b}{(2 n + 1) \pi}
  -
  \frac{8\big(T_b- T_a\big) }{(2n+1)^2 \pi^2} 
  \ .
\end{eqnarray}
\subsubsection{BC4}

For the boundary conditions
\begin{eqnarray}
  \partial_x T(0,t) &=& F_1
  \\[5pt]
  T(L,t) &=& T_2
  \ ,
\end{eqnarray}
the full nonhomogeneous solution is
\begin{eqnarray}
  \bar T(x) &=& 
  T_1 + \frac{T_2 - T_1}{L} \, x
  +
  \sum_{n=0}^\infty 
  A_n \, \cos k_n x\, e^{-\kappa \, k_n^2 t}
  \\[5pt]
  T_1 &=& T_2 - F_1 L
  =
  \frac{\gamma_2}{\alpha_2} - \frac{\gamma_1 L}{\beta_1}
  \\[5pt]
    k_n &=& 
  \frac{(2 n + 1) \pi}{2 L}
  \hskip1.0cm 
  n = 0, 1, 2, \cdots
  \ .
\end{eqnarray}
As before, we take the linear initial condition (\ref{IClinear}),
and then (\ref{eq_Tcoeff}) gives the coefficients
\begin{eqnarray}
  A_n 
  &=&
  4 T_a\, \frac{(-1)^n }{(2 n + 1)\pi }
  -
  8\Big(T_b - T_a \Big) \,\frac{1 - (-1)^n}{(2n + 1)^2\, \pi^2}
  \ .
\end{eqnarray}
\begin{figure}[h!]
\includegraphics[scale=0.45]{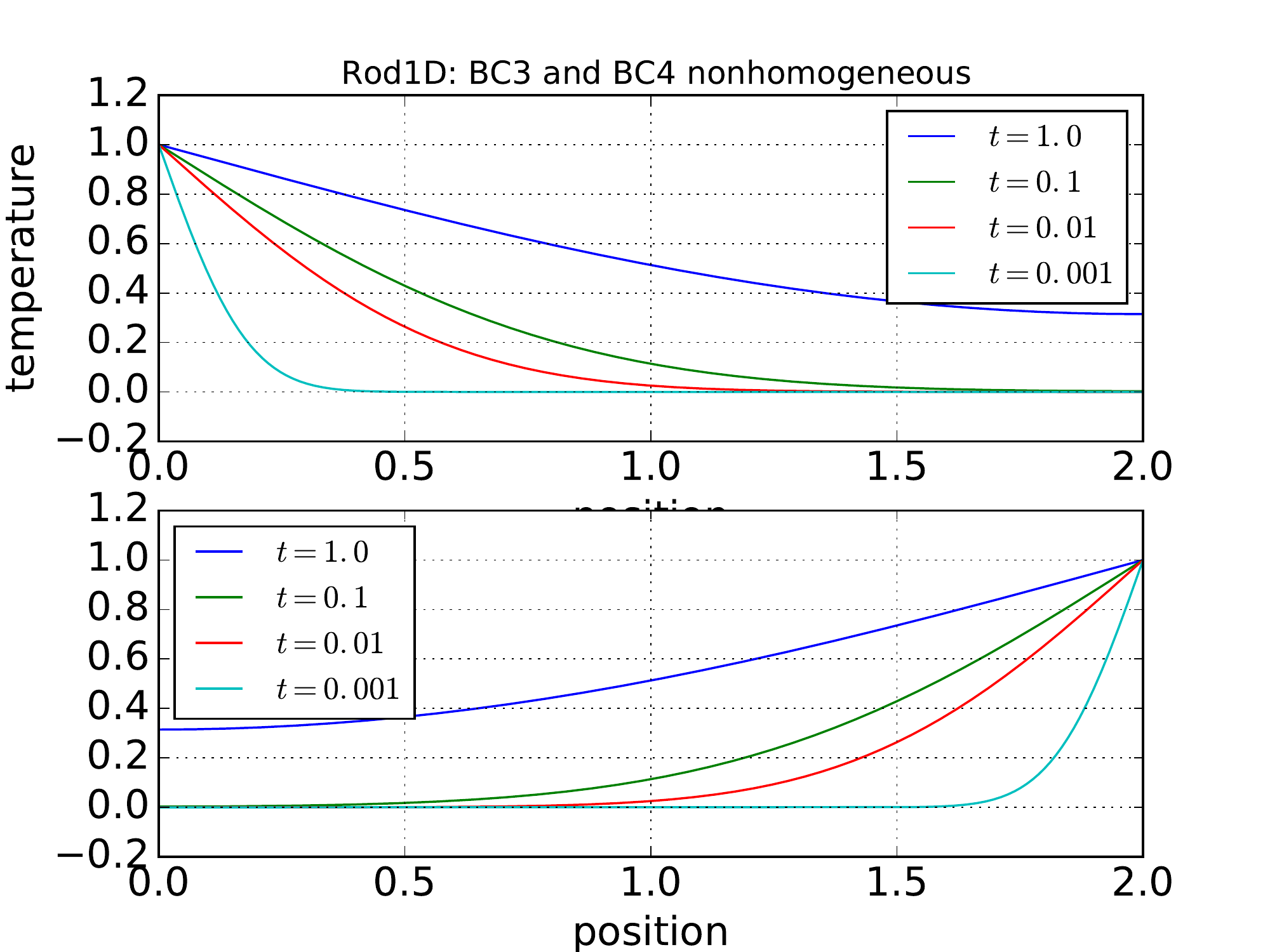} 
\caption{\footnoteskip  
  BC3  and BC4 for $\kappa=1$, $L=2$, $T_1 = 1, T_2=0$,
  $T_\smL=T_\smR=0$. 
  The two profiles should be mirror images of each other, 
  by symmetry principle. This appears to be the case, for  for 
  $N_{\rm max}=300$. Note that the profile are indeed asymmetric.
  BC3: Rod1D(alpha1=1, beta1=0, alpha2=0, beta2=1, TL=3, TR=4). 
  BC4:  Rod1D(alpha1=0, beta1=1, alpha2=1, beta2=0, TL=4, TR=3).
}
\label{fig_rod1D_BC3}
\end{figure}

\subsection{General Boundary Conditions}

For general boundary conditions, the full nonhomogeneous solution is
\begin{eqnarray}
  T(x,t)
  &=& 
  T_1 + \frac{T_2 - T_1}{L}\, x
  +
  \sum_{n=1}^\infty D_n \, X_n(x) \, e^{-\kappa \, k_n^2 t}
  \\[5pt]
  X_n(x)
  &=& 
   A_n \cos k_n x + B_n \sin k_n x 
  \ ,
\end{eqnarray}
with coefficients
\begin{eqnarray}
  A_n &=& - \frac{\beta_1 k_n}{\alpha_1}\, B_n
  \\[5pt]
  T_1
  &=& 
  \frac{\beta_2 \gamma_1 -\beta_1 \gamma_2 + L \alpha_2 \gamma_1}
  {\alpha_1 \beta_2 - \alpha_2 \beta_1 + L \alpha_1 \alpha_2 }
  \\[5pt] 
  T_2
  &=&  
  \frac{\beta_2 \gamma_1 -\beta_1 \gamma_2 + L \alpha_1 \gamma_2}
  {\alpha_1 \beta_2 - \alpha_2 \beta_1 + L \alpha_1 \alpha_2 }
\ .
\end{eqnarray}
The Fourier coefficients are
\begin{eqnarray}
    D_n &=& \frac{1}{N_n}\int_0^L dx\, \Big[T_0(x) - \bar T(x)\Big] X_n(x)
  \ .
\end{eqnarray}
The zeroth order contributions is $T_0^{(0)}(x) - \bar T^{(0)}(x)=T_a$,
and we find
\begin{eqnarray}
  D_n^{(0)} &=& \frac{T_a}{N_n} \, \left[
  \frac{1 - \cos k_n L}{k_n}  - \frac{\beta_1 \sin k_n L}{\alpha_1}\right]
  \ .
\end{eqnarray}
The first order contribution is $T_0^{(1)}(x)- \bar T^{(0)}(x) = 
(T_b - T_a) \, x/L$ we have
\begin{eqnarray}
  D_n^{(1)} &=& \frac{T_b - T_a}{N_n\, L}\,\frac{1}{\alpha_1 k_n^2} \, \Big[
  \beta_1 k_n  - (\alpha_1 k_n L + \beta_1 k_n) \cos k_n L + (\alpha_1 - 
  \beta_1 k_n^2 L) \sin k_n L 
  \Big]
  \ .
\end{eqnarray}
The normalization factor is
\begin{eqnarray}
  &&N_n
  =
  \frac{1}{4 k_n \alpha_1^2}\Big[
  -2 \alpha_1 \beta_1 k_n  + 2 (\beta_1^2 k_n^2 + \alpha_1^2 ) k_nL +
  2\alpha_1 \beta_1 k_n \cos 2 k_n L + 
  (\beta_1^2 k_n^2 - \alpha_1^2 ) \sin 2 k_n L
  \Big]
  \nonumber
  \\
  \ . 
  \label{SolABfour}
\end{eqnarray}
Setting $\mu \equiv k L$ and $\bar\beta_i \equiv \beta_i/L$, we can write (\ref{BCtankLone}) in the form
\begin{eqnarray}
   \tan \mu
   &=&
   \frac{ (\alpha_2 \bar\beta_1 - \alpha_1 \bar\beta_2) \, \mu }{\alpha_1 \alpha_2  +  \bar\beta_1 \bar\beta_2\, \mu^2  }   
     \label{BCtanmuAgain}   
   \ .
\end{eqnarray}
Equation (\ref{BCtanmuAgain}) will give solutions $\mu_n$ for $n=0,1,2,\cdots$ (with $\mu_0=0)$,
and the wave numbers become
\begin{eqnarray}
  k_n = \frac{\mu_n}{L}
  \ .
  \label{knDef}
\end{eqnarray}
\begin{acknowledgments}
 I would like to thank Jim Ferguson and Scott Doebling for carefully
 reading through the text. 
\end{acknowledgments}

%
\appendix

\section{Sample ExactPack Script}
\label{sec:sample_ep_script}

The following script produces Fig.~\ref{fig_planar_sandwich_ep}. 

\footnoteskip
\begin{verbatim}
import numpy as np
import matplotlib.pylab as plt

from exactpack.solvers.heat import PlanarSandwich

L = 2.0
x = np.linspace(0.0, L, 1000)
t0 = 1.0
t1 = 0.2
t2 = 0.1
t3 = 0.01
t4 = 0.001

solver = PlanarSandwich(T1=1, T2=0, L=L, Nsum=1000)
soln0 = solver(x, t0)
soln1 = solver(x, t1)
soln2 = solver(x, t2)
soln3 = solver(x, t3)
soln4 = solver(x, t4)
soln0.plot('temperature', label=r'$t=1.000$')
soln1.plot('temperature', label=r'$t=0.200$')
soln2.plot('temperature', label=r'$t=0.100$')
soln3.plot('temperature', label=r'$t=0.010$')
soln4.plot('temperature', label=r'$t=0.001$')

plt.title('Planar Sandwich')
plt.ylim(0,1)
plt.xlim(0,L)
plt.legend(loc=0)
plt.grid(True)
plt.show()
\end{verbatim}
\bodyskip

\section{Uniformly Convergent Sequences of Functions}

Many of the mathematical operations we take for granted 
in a typical analytic calculation of a physical process, such 
as the {\em simple} interchange of a limit and an integral, 
depend deeply upon issues surrounding the uniform convergence 
of sequences of functions. By way of introduction, let us consider
a solution $T(x,t)$ to the heat flow equations
(\ref{eq_oneDrodAnh})--(\ref {eq_oneDrodCnh}). Let us further
consider a sequence of times $t_1, t_2, t_3, \cdots$, from which
we can construct a sequence 
of temperature profiles $T_n(x) = T(x,t_n)$. In other words, $T_n(x)$ 
is a sequence of functions of $x$, indexed by the integers $n$, or 
equivalently by the times $t_n$. Suppose now that the time sequence 
$t_n$ converges to the limit $t_0$, so that  $\lim_{n \to \infty} 
t_n = t_0$.  Then, for our purposes, we may speak interchangeably 
of the limits $\lim_{n \to\infty} T_n(x)$ and $\lim_{t \to t_0} T(x,t)$, 
and in this way, we can think of $T(x,t)$ as a sequence of functions 
of $x$ indexed by $t$.  To make this more precise, and to refresh 
our memories, it is constructive to review the formal definition of a 
limit. The sequence $\{ t_n\}$ converges to the the limit $t_0$ as 
$n \to \infty$, denoted
\begin{eqnarray}
  \lim_{n \to \infty} t_n = t_0
  \ ,
\end{eqnarray}
provided that for every $\epsilon > 0$ there exists $N > 0$
such that 
\begin{eqnarray}
  \big\vert t_n(x) - t_0 \big\vert < \epsilon
\end{eqnarray}
whenever $n \ge N$. That is to say, $t_n$ can be made arbitrarily
close to $t_0$ by choosing $n$ arbitrarily large.

The notion of a limit can extended to a sequence of functions. The 
domain of the functions $T_n(x)$, which we refer to as $E$, can be 
either the open interval $(0,L)$, or the closed interval $[0,L]$, if we 
are also interested in the boundary points $x=0,L$.  For definiteness, 
we take the case BC1, for which $T(0,t_n)=T_1$ and $T(L,t_n)=T_2$. 
There are two distinct (but related) sense in which the limit
\begin{eqnarray}
  \lim_{n \to \infty} T_n(x) = T(x)
\end{eqnarray}
exists.
The obvious way to interpret this limit is to choose a value of
$x=x_0$, and to take the limit of the normal sequence of numbers
$T_1(x_0), T_2(x_0), T_3(x_0), \cdots$. If, in the limit $n \to \infty$,
the sequence converges to a number $T(x_0)$ for some function
$T(x)$, we say that the sequence $T_n(x)$ converges point-wise
to $T(x)$ at $x=x_0$. This is made formal by the following 
definition.

\vskip0.3cm
\noindent
{\bf Definition}: The sequence of functions $\{T_n(x)\}$
converges {\em point-wise} on $E$ to a function $T(x)$ if
for every $x \in E$ and for every $\epsilon>0$ there is an integer 
$N$ such that
\begin{eqnarray}
\label{fnfepsilon}
  \big\vert T_n(x) - T(x) \big\vert < \epsilon
\end{eqnarray}
for all $n \ge N$. 

\vskip0.2cm
\noindent
The integer $N$ might depend upon the point $x$. If, however,
we can choose the same $N$ for all $x \in E$, then we say that
the limit is {\em uniformly} convergent. This is made precise in
following definition.

\begin{figure}[t!]
\includegraphics[scale=0.40]{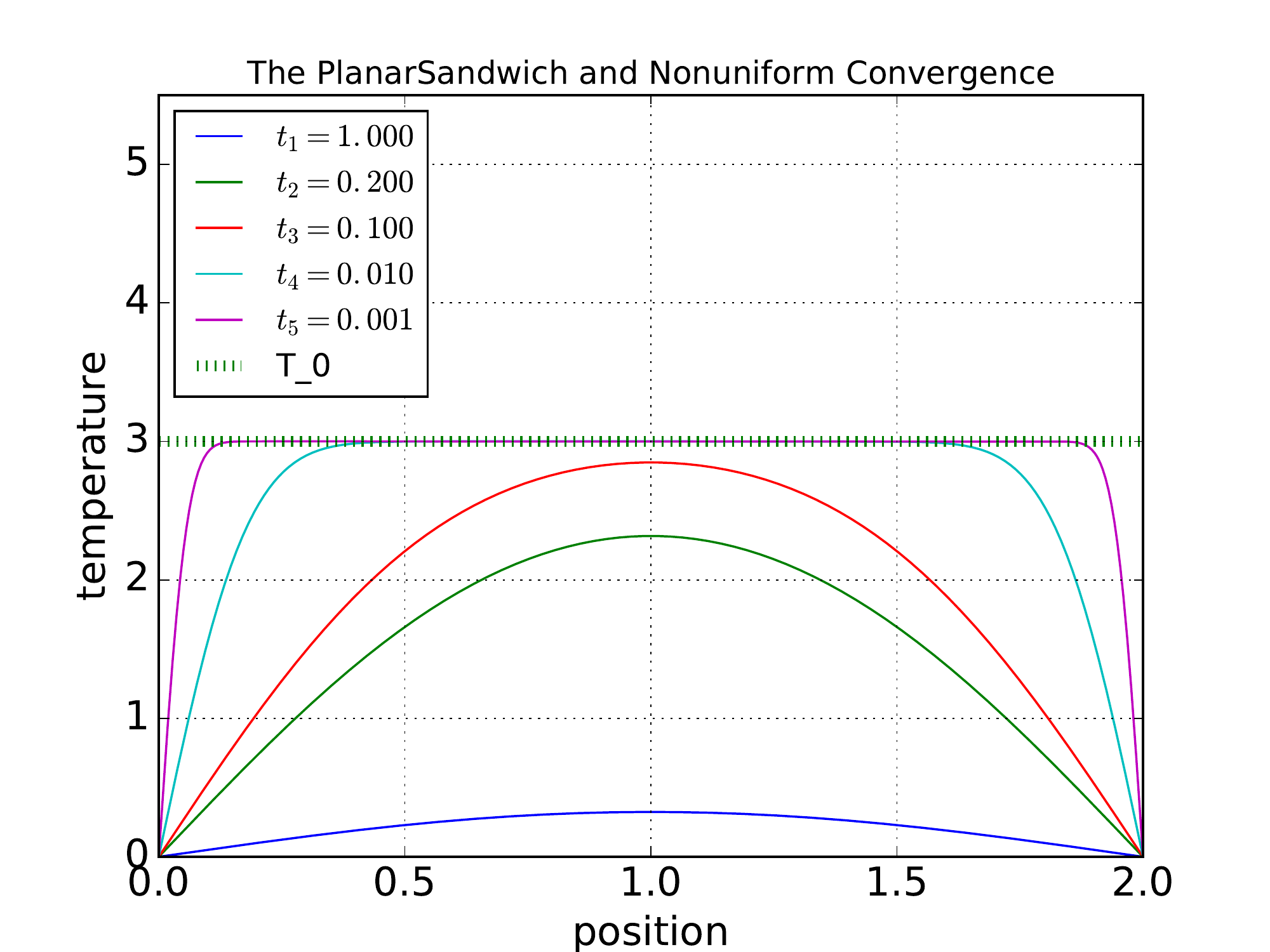} 
\caption{\footnoteskip  
Temperature profiles for the homogeneous planar sandwich at times 
$t_1=1$, $t_2=0.2$, $t_3=0.1$, $t_4=0.01$, and $t_5=0.001$.
The diffusion constant is $\kappa=1$ and length of the rod is
$L=2$, with a constant initial condition $T_0(x)=3$. The plot uses 
the instance PlanarSandwich(T1=0, T2=0, TL=3, TR=3, L=2, 
Nsum=1000). Since the boundary conditions are incommensurate
with the initial condition,  the solution $T(y,t)$ convergens 
non-uniformly on the open $x$-interval $(0,L)$ to $T_0(x)=3$,
which is plotted by the dashed line.
}
\label{fig_planar_sandwich_nonuniform_ep}
\end{figure}

\vskip0.3cm
\noindent
{\bf Definition}: The sequence of functions $\{T_n(x)\}$
converges {\em uniformly} on $E$ to a function $T(x)$ if
for every $\epsilon>0$ there is an integer $N$ such that
\begin{eqnarray}
\label{fnfepsilon}
  \big\vert T_n(x) - T(x) \big\vert < \epsilon
\end{eqnarray}
for all $n  \ge N$ and all $x \in E$. 
\vskip0.3cm

\noindent
As an example, let us consider the solution illustrated in 
Fig.~\ref{fig_planar_sandwich_nonuniform_ep}. This is a homogeneous
solution, for which $T(0,t)=T(0,L)=0$, with a constant initial
condition $T_0(x)=3$ (for $0 < x < L$). The time sequence
is $t_1=1$, $t_2=0.2$, $t_3=0.1$, $t_4=0.01$, $t_5=0.001, 
\cdots$. We see that $\lim_{n \to \infty}T_n(x) = T_0(x)$ for
$x \in (0,L)$, but the limit is non-uniform.

\vfill

\end{document}